\documentclass[aps,prb,preprint]{revtex4}
\usepackage[dvipdfm]{graphicx} 
\usepackage{bm}
\usepackage{booktabs}
\usepackage{amsmath,amsthm,amssymb}
\usepackage{ascmac}
\usepackage{comment}

\makeatletter
\AtBeginDocument{\@ifpackageloaded{natbib}{\ifNAT@numbers\if@filesw\immediate\write\@auxout{\string\global\string\NAT@numberstrue}\fi\fi}{}}
\makeatother
\begin{document}

\title{Spin and charge transport induced by a twisted light beam on a surface of a topological insulator}
\author{Kunitaka Shintani, Katsuhisa Taguchi, Yukio Tanaka, and, Yuki Kawaguchi }

\affiliation{Department of Applied Physics, Nagoya University, Nagoya, 464-8603, Japan}
\date{\today}
%
\begin {abstract}
We theoretically study spin and charge transport induced by a twisted light beam irradiated on a disordered surface of a doped three dimensional topological insulator (TI). 
We find that various types of spin vortices are imprinted on the surface of the TI depending on the spin and orbital angular momentum of the incident light. 
The key mechanism for the appearance of the unconventional spin structure is the spin-momentum locking in the surface state of the TI.
Besides, the diffusive transport of electrons under an inhomogeneous electric field causes a gradient of the charge density, which then induces nonlocal charge current and spin density as well as the spin current. 
We discuss the relation between these quantities within the linear response to the applied electric field using the Keldysh-Green's function method.

\begin{description}
\item[] \hspace{9.48cm}  PACS numbers: 78.20.Ls
\end{description}
\end{abstract}

\maketitle
\section{Introduction}\label{sec:i}
Emergence and manipulation of spins are a major research topic in spintronics. 
Applying a controlled light is one of the promising techniques to manipulate spins. 
Recently, the spin angular momentum of a circularly polarized light has been observed to induce the magnetization in solid-state materials through spin-orbit interactions\cite{rf:Kimel05,rf:Hansteen05,rf:Kirilyuk,rf:Iida11}.
This technique has further been applied to the ultrafast magnetization switching, whose time is much shorter than that by an applied magnetic field\cite{rf:Stanciu07,rf:Vahaplar09}.

When a light is irradiated on a surface of a three-dimensional topological insulator (TI), spin is predicted to emerge in the perpendicular direction to the electric field of the light\cite{rf:Raghu10,rf:Misawa11,rf:Liu13}. 
Here, a TI is an anomalous material with strong spin-orbit interactions.
Electrons insulate in the bulk, while they conduct on the surface of the TI, where 
the exotic surface state of the TI is caused by both the spin-orbit interaction and the topological electric structure\cite{rf:Hasan10,rf:Qi11,rf:Ando13}.
On the surface of the TI, the direction of the spin and that of the momentum are perfectly locked to be perpendicular to each other, which is dubbed spin-momentum locking. 
Because of this spin-momentum locking, the charge current generated along the direction of the electric field causes the spin density in the perpendicular direction\cite{rf:Raghu10,rf:Misawa11,rf:Liu13}.
Such a manipulation of spin and charge current using a light may make it possible to develop magneto-optical devices based on TIs\cite{rf:Raghu10,rf:Misawa11,rf:Liu13,rf:Tse10,rf:Tse10R}. 

Recently, magneto-optical effects and optical excitation using a twisted light beam, whose phase is twisted around the direction of the propagation of light, have been theoretically predicted\cite{rf:Quinteiro09,rf:Quinteiro10,rf:Quinteiro11,rf:Watzel12,rf:Tamborenea13,rf:Tamborenea15} and experimentally carried out\cite{rf:Ueno09,rf:Clayburn13}. 
A twisted light has the following two intriguing properties distinct from a plane wave\cite{rf:Marrucci}.
First, the phase of the light is twisted around the center of the beam, and hence, has a singularity at the center. 
As a result, strength of the light becomes zero at the center of the beam.
Second, because of the twisted phase, the strength of the light strongly depends on the space, whose distributions are manipulated by the angular momentum of the light. 
The above properties can be well-understood by writing down the electric field of the light.
The electric field of the twisted light beam traveling along the $z$ axis at $z=z_0$, $\bm{E}=(E_x,E_y)$, can be described by\cite{rf:Allen,rf:Molina,rf:Marrucci,rf:Padgett04,rf:Mendonca09,rf:Lembessis09} 
\begin{align}\label{eq:1-1} 
\bm{E} (r,\varphi, t, z_0)& = \mathcal{E}(r, z_0) \textrm{Re} [ (1, i \sigma_\textrm{L}^z) e^{i(q_z z_0 - \Omega t )} e^{i m^z_{\textrm{L}} \varphi} ],
\end{align}
where $(r,\varphi)$ is the two-dimensional polar coordinates at $z=z_0$, 
$t$ is the time, and $q_z$ and $\Omega$ are the momentum and frequency of the twisted light beam, respectively. 
Here, $\mathcal{E}(r,z_0)$ denotes the magnitude of the electric field, which depends on the space and becomes zero at the center $r=0$ for a nonzero $m^z_{\textrm{L}}$ due to the phase singularity.
$\sigma^z_{\textrm{L}}=1, -1$ and $m^z_{\textrm{L}}=0, \pm1, \pm2, \cdots$ represent the $z$ components of the spin and orbital angular momentum of the light, respectively. 
The former corresponds to the direction of the circular polarization, i.e., $\sigma^z_{\textrm{L}}=1(-1)$ represents a right-handed (left-handed) circularly polarized wave, while the latter describes the winding of the electric field in the $z=z_0$ plane. 
In fact, the electric field of a twisted light has the topological quantity. 
We will see later that the winding number of a twisted light given by Eq. (\ref{eq:1-1}) is proportional to $\sigma_{\textrm{L}}^z$ and $m_{\textrm{L}}^z$[see the discussion below Eq. (\ref{eq:winding-number})].

So far, it has been theoretically predicted that in the presence of the spin-orbit interaction, unconventional photo-induced spin excitation and current emerge due to the spatial dependence of the strength of the electric field of the twisted light. 
However, the orbital angular momentum of the light is not transfered to the spin polarization since the beam waist is much larger than the width of the electron wave function, which is in the order of the lattice constant, and hence each electrons feels locally uniform electric field. 
Actually the experimental investigation of the photo-induced spin polarization\cite{rf:Clayburn13} could not detect the orbital angular momentum dependence in the semiconductor with the Rashba and Dresselhaus type spin orbit interaction as an exception there is a theoretical prediction for cylindrical quantum disks\cite{rf:Quinteiro09}.


%
%
%
%
%

%
%
In this paper, we theoretically study spin and charge generation due to the electric field of the twisted light beam on a disordered surface of a doped TI by using the Green's function technique.
We analytically calculate the linear response function of the spin density to a space-time dependent external electric field. 
We find that the local and nonlocal spin densities are induced by the electric field and the gradient of the electric field, respectively, via the spin-momentum locking. 
Here, the local spin density comes from the charge current that flows along the electric field, whereas the nonlocal one couples to the diffusive charge current due to the impurity scatterings on the disordered surface of the TI. 
In addition, the gradient of the electric field also induces the charge density and the spin current.
Applying the obtained results to the electric field of a twisted light beam, 
we find that various spin distributions appear depending on the orbital as well as spin angular momentum of the light. 
Moreover the spin distributions have topological structures i.e., magnetic vortex-like textures, characterized with winding numbers, which dependent on both $\sigma^z_{\textrm{L}}$ and $m^z_{\textrm{L}}$.
The induced spin structure evolves in time but its winding number remains a constant. 
The manipulation of such a topological spin structure could be applicable for the spintronics related to magnetic vortices and skyrmions.

%
%
This paper is organized as follows. 
In Sec. \ref{sec:ii}, we introduce the model Hamiltonian for the disordered surface of the TI in the presence of a space-time dependent electromagnetic field. 
We also present the Green's functions on the disordered surface of the TI.
In Sec. \ref{sec:iii}, 
we calculate the induced spin density on the surface of the TI within the linear response to the applied electric field.
Section IV discusses the general properties of the charge density, spin density, charge current, and spin current induced by the electric field on the surface of the TI.
Section \ref{sec:4} discusses the properties of the twisted-light-induced spin and charge distributions.
Section \ref{sec:5} summarizes the paper.
Appendices \ref{sec:A}-\ref{sec:G} give the detailed calculations used in Sec. III.

\section{Model}\label{sec:ii}
We consider a disordered surface of a three-dimensional TI and apply a space-time dependent electromagnetic field to it.
We assume that impurities on the disordered surface are nonmagnetic.
The Hamiltonian we consider is given by  
\begin{align}\label{eq:2-1} 
\mathcal{H} & = \mathcal{H}_{\textrm{TI}} + \mathcal{H}_{\textrm{em}} +\mathcal{V}_{\textrm{imp}}, 
	\\  \label{eq:2-2} 
\mathcal{H}_{\textrm{TI}} & =  \int d\bm{x} \psi^\dagger [ -i\hbar v_{\rm{F}} (\hat{ \bm{\sigma}} \times \bm{\nabla})_z - \epsilon_{\textrm{F}} ]\psi, 
	\\ \label{eq:2-3} 
\mathcal{H}_{\textrm{em}} & = - ev_{\textrm{F}} \int d\bm{x} \psi^\dagger  (\hat{ \bm{\sigma}} \times \bm{A}_{\textrm{em}})_z \psi,
	\\ \label{eq:2-4} 
\mathcal{V}_{\textrm{imp}} & =  \int d\bm{x} \ u_\textrm{i} \psi^\dagger \psi,   
\end{align} 
where $\psi^\dagger \equiv \psi^\dagger(\bm{x},t) = (\psi^\dagger_{\uparrow} \ \psi^\dagger_{\downarrow})$ and $\psi$ are the creation and annihilation operators of conduction electrons on the surface of the TI, $\hat{\sigma}_{j(=x,y,z)}$ are the Pauli matrices, and $e<0$ is the elementary charge of electrons.
Here, we assume a doped TI, and $\epsilon_{\rm{F}}$ and $v_{\rm{F}}$ are the Fermi energy and the Fermi velocity, respectively, on the surface of the doped TI. 
$\mathcal{H}_{\textrm{em}}$ is the gauge coupling between conduction electrons and the electromagnetic field. 
The vector potential of the electromagnetic field $\bm{A}_{\textrm{em}}$ generally depends on the space and time, and the electric field and the magnetic field are respectively given by $\bm{E}=-\partial_t \bm{A}_{\textrm{em}}$ and $\bm{B}=\bm{\nabla}\times \bm{A}_{\textrm{em}}$.
$\mathcal{V}_{\textrm{imp}}$ describes the potential due to the nonmagnetic impurity scatterings, 
where $u_\textrm{i}(\bm{x}) = \sum_{j=1}^{N_\textrm{i}} u_0\left( \delta(\bm{x}-\bm{R}_{j})-\frac{1}{L^2}\right) $ 
is the potential energy density with $N_\textrm{i}$ being the number of the impurities, $u_0$ a constant, $\bm{R}_{j}$ the position of the $j$-th impurity on the surface, and $L^2$ the area of the surface. 
Here, the contribution from the impurity potential is treated as the impurity average $\langle u_\textrm{i}(\bm{q}) u_\textrm{i}(\bm{q}')  \rangle_\textrm{i} = \frac{N_i u_0^2}{L^4} \delta_{q,q'}$, where $u_\textrm{i}(\bm{q})$ is the Fourier transform of $u_\textrm{i}(\bm{x})$.

To calculate the spin density and the charge density, we use the Green's function method. 
In the absence of the electromagnetic field, the retarded Green's function is given by \cite{rf:book2,rf:Fujimoto13,rf:sakai14} 
\begin{align} \label{eq:2-5} 
\hat{g}^{\textrm{r}}_{\bm{k},\omega} 
	 = \left[ \hbar \omega +\epsilon_{\textrm{F}} -\hbar v_{\textrm{F}} \hat{ \bm{\sigma}} \cdot (  \bm{k} \times \bm{z})  - \hat{\Sigma}_{\bm{k},\omega} \right]^{-1}.
\end{align}
Here, a variable with a hat denotes a two-by-two matrix. 
By calculating the self-energy $\hat{\Sigma}_{\bm{k},\omega}$ within the self-consistent Born approximation \cite{rf:book2,rf:Fujimoto13,rf:sakai14} and expanding it with respect $\bm{k}$ up to the linear terms,  \cite{rf:Fujimoto13,rf:sakai14,rf:Taguchi15} 
Eq. (\ref{eq:2-5}) is rewritten as
\begin{align} \label{eq:2-6} 
\hat{g}^{\textrm{r}}_{\bm{k},\omega} 
	 = \left[ \hbar \omega +\epsilon_{\textrm{F}} -\hbar \tilde{v}_{\textrm{F}} \hat{ \bm{\sigma}} \cdot (  \bm{k} \times \bm{z})  + i\eta \right]^{-1},
\end{align}
where $\tilde{v}_{\textrm{F}}=v_{\textrm{F}}/(1+\xi)$ is the modified Fermi velocity due to nonmagnetic impurity scatterings with 
$\xi=n_{\textrm{i}} u_0^2/(4\pi\hbar^2v_{\textrm{F}}^2)$ being a constant depending on the properties of the TI, 
and the imaginary part of the self-energy $\eta = \pi n_\textrm{i} u_0^2 \nu_e/2$ defines the transport relaxation time $\tau= \hbar /(2\eta)$.
Here, $n_\textrm{i} = N_\textrm{i}/L^2$ is the concentration of the impurities on the surface and $\nu_e$ is the density of states of electrons on the surface. 
Since we are considering a metallic state, $\tau$ satisfies $\hbar/(\epsilon_{\textrm{F}} \tau)\ll 1$. 
By comparing Eqs. (\ref{eq:2-2}) and (\ref{eq:2-6}),
we see that the effective Hamiltonian for the surface electrons affected by impurities is given by the right-hand side of Eq. (\ref{eq:2-2}) with replacing $v_{\textrm{F}}$ with $\tilde{v}_{\textrm{F}}$.
Accordingly, $v_{\textrm{F}}$ in Eq. (\ref{eq:2-3}) is replaced by $\tilde{v}_{\textrm{F}}$.

\section{Spin and charge densities induced by an applied electric field}\label{sec:iii}
\label{sec:3} 
\begin{figure}[htbp]\centering 
\includegraphics[scale=0.3]{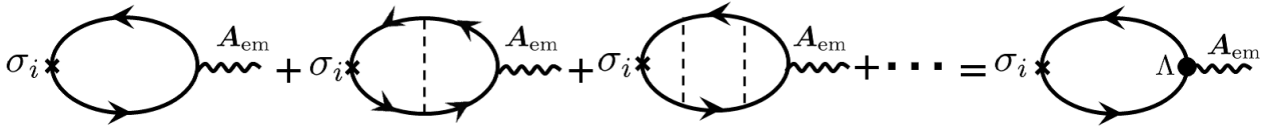}
\caption{Vertex function due to $\mathcal{V}_{\textrm{imp}}$. The dashed and wavy lines mean the potential due to the nonmagnetic impurity scatterings and gauge coupling between conduction electrons and the electromagnetic field, respectively.
}\label{fig:diagram}
\end{figure}

We calculate the spin density induced by an applied electric field on a disordered surface of a doped TI by using the Keldysh Green's function method within the linear response to $\mathcal{H}_\textrm{em}$.
The spin density $\bm{s} = \frac{1}{2}\langle \psi^\dagger \hat{\bm{\sigma}} \psi \rangle $ is described by using the lesser component of the Keldysh-Green's function in the same position and time $-i\hbar \hat{G}^{<} (\bm{x},t, \bm{x},t) = \langle \psi^\dagger(\bm{x},t) \psi(\bm{x},t) \rangle$ as 
\begin{align}\label{eq:3-1} 
s_{i} (\bm{x},t) & =  -\frac{i\hbar}{2}  {\textrm{tr}} \bigl[\hat{\sigma}_i  \hat{G}^{<} (\bm{x},t, \bm{x},t)\bigr] \hspace{5mm}(i=x,y,z),
\end{align} 
where ${\textrm{tr}}$ denotes the trace over the spin indices. 
Then, from the Dyson's equation for $ \hat{G}^{<} (\bm{x},t, \bm{x},t)$ the induced spin density within the linear response to $\mathcal{H}_\textrm{em}$ is given by 
\begin{align}\label{eq:3-2} 
 s_{\mu}(\bm{x},t)	& =  \frac{i \hbar e \tilde{v}_{\textrm{F}} }{2L^2} 
 		\sum_{\nu, u = x, y, z }\epsilon_{z \nu u}
		 \sum_{\bm{q},\Omega}e^{i(\Omega t-\bm{q}\cdot\bm{x})} 
		{\textrm{tr}}[ \hat{\Pi}_{\mu \nu} (\bm{q},\Omega) ] A_{{\textrm{em}}, u} ({\bm{q},\Omega}), 
\end{align}
where $\epsilon_{z  \nu u}$ is the Levi-Civita symbol,
$\hat{\Pi}_{\mu \nu}$ is the spin-spin response function, 
and $\bm{q}=(q_x, q_y)$ and $\Omega$ are the momentum and frequency of $A_{{\textrm{em}}, u} ({\bm{q},\Omega})$, respectively.
The response function $\hat{\Pi}_{\mu \nu}$ can be decomposed as $\hat{\Pi}_{\mu \nu} = \hat{\sigma}_\mu \hat{\Pi}_{\nu} $ and $\hat{\Pi}_{\nu}$ is represented by 
\begin{align} \label{eq:3-3} 
\hat{\Pi}_{\nu}(\bm{q},\Omega)  
	& = \sum_{\bm{k}, \omega} [ \hat{g}_{\bm{k}-\frac{q}{2},\omega-\frac{\Omega}{2}} \hat{\Lambda}_\nu (\omega,\bm{q},\Omega) \hat{g}_{\bm{k}+\frac{q}{2},\omega+\frac{\Omega}{2}} ]^{<}.
\end{align}
Here, $\hat{\Lambda}_\nu$ is the vertex function due to $\mathcal{V}_{\textrm{imp}}$, whose diagram is shown in the Fig. \ref{fig:diagram}, and is given by 
\begin{align} \label{eq:3-4} 
	\hat{\Lambda}_\nu(\omega,\bm{q},\Omega)
		& =  \hat{\sigma}_\nu + \sum_{\mu=0,x,y,z}\sum_{n=1}^\infty {[\Gamma(\omega,\bm{q},\Omega) ]^n}_{\nu \mu}\hat{\sigma}_\mu,
\end{align}
where $\hat{\sigma}_0$ is the two-by-two identity matrix and $\Gamma$ is a $4\times 4$ matrix defined from the following equation 
\begin{align}\label{eq:3-5} 
	\hat{ \Gamma }_\nu(\omega,\bm{q},\Omega)
		&\equiv n_{\textrm{i}} u_{\textrm{i}}^2  \sum_{\bm{k}}  \hat{g}_{\bm{k}-\frac{q}{2},\omega-\frac{\Omega}{2}} \hat{\sigma}_\nu \hat{g}_{\bm{k}+\frac{q}{2},\omega+\frac{\Omega}{2}}
	\\
		&=\sum_{\mu=0,x,y,z}\Gamma_{\nu \mu}(\omega,\bm{q},\Omega)\hat{\sigma}_\mu.
\end{align}
 
Expanding Eq. (\ref{eq:3-3}) with respect to the retarded and advanced Green's functions, $\hat{g}^{\textrm{r}}$ and $\hat{g}^{\textrm{a}}$, based on the formula\cite{rf:book1} $\hat{g}^<_{\bm{k},\omega} = f_\omega (\hat{g}^\textrm{a}_{\bm{k},\omega} - \hat{g}^\textrm{r}_{\bm{k},\omega} )$, where $f_\omega \equiv 1/(e^{\beta\hbar \omega}+1)$ is the Fermi distribution function, the spin-spin response function can be divided into three terms:
\begin{align} \label{eq:3-6} 
\hat{\Pi}_{\nu}(\bm{q},\Omega)  
	& =  \hat{\Pi}^{\textrm{ra}}_{\nu}(\bm{q},\Omega) + \hat{\Pi}^{\textrm{rr}}_{\nu}(\bm{q},\Omega) + \hat{\Pi}^{\textrm{aa}}_{\nu}(\bm{q},\Omega), 
	\\  \label{eq:3-7} 
\hat{\Pi}^{\textrm{ra}}_{\nu}(\bm{q},\Omega) 
	& \equiv  \sum_{\bm{k}, \omega} (f_{\omega+\frac{\Omega}{2}} -f_{\omega-\frac{\Omega}{2}} )   \hat{g}^\textrm{r}_{\bm{k}-\frac{\bm{q}}{2},\omega-\frac{\Omega}{2}} \hat{\Lambda}^{\textrm{ra}}_\nu(\omega,\bm{q},\Omega)  \hat{g}^\textrm{a}_{\bm{k}+\frac{\bm{q}}{2},\omega+\frac{\Omega}{2}},
	\\ \label{eq:3-8} 
\hat{\Pi}^{\textrm{rr}}_{\nu}(\bm{q},\Omega) 
	& \equiv - \sum_{\bm{k}, \omega} f_{\omega+\frac{\Omega}{2}}   \hat{g}^\textrm{r}_{\bm{k}-\frac{\bm{q}}{2},\omega-\frac{\Omega}{2}} \hat{\Lambda}^{\textrm{rr}}_\nu(\omega,\bm{q},\Omega)  \hat{g}^\textrm{r}_{\bm{k}+\frac{\bm{q}}{2},\omega+\frac{\Omega}{2}} ,
	\\ \label{eq:3-9} 
\hat{\Pi}^{\textrm{aa}}_{\nu}(\bm{q},\Omega) 
	& \equiv  \sum_{\bm{k}, \omega} f_{\omega-\frac{\Omega}{2}}    \hat{g}^\textrm{a}_{\bm{k}-\frac{\bm{q}}{2},\omega-\frac{\Omega}{2}} \hat{\Lambda}^{\textrm{aa}}_\nu(\omega,\bm{q},\Omega)  \hat{g}^\textrm{a}_{\bm{k}+\frac{\bm{q}}{2},\omega+\frac{\Omega}{2}}.
\end{align}
Here, $\hat{\Lambda}^{\textrm{AB}}_\nu$ (\textrm{A,B} = \textrm{r,a}) is defined by
\begin{align} \label{eq:3-4-2} 
	\hat{\Lambda}^{\textrm{AB}}_\nu(\omega,\bm{q},\Omega)
		& =  \hat{\sigma}_\nu + \sum_{n=1}^\infty {[ {\Gamma}^{\textrm{AB}}(\omega,\bm{q},\Omega) ]^n}_{\nu \mu}\hat{\sigma}_{\mu},
	\\ \label{eq:3-5-2} 
	\hat{ \Gamma }^{\textrm{AB}}_\nu(\omega,\bm{q},\Omega)
		&\equiv n_{\textrm{i}} u_{\textrm{i}}^2  \sum_{\bm{k}}  \hat{g}^{\textrm{A}}_{\bm{k}-\frac{q}{2},\omega-\frac{\Omega}{2}} \hat{\sigma}_\nu \hat{g}^{\textrm{B}}_{\bm{k}+\frac{q}{2},\omega+\frac{\Omega}{2}},
		\\
		&=\sum_{\mu=0,x,y,z}\Gamma^{\textrm{AB}}_{\nu \mu}(\omega,\bm{q},\Omega)\hat{\sigma}_\mu.
\end{align}
By expanding
$\hat{\Gamma}^{\textrm{rr}}_{\nu}$ and $\hat{\Gamma}^{\textrm{aa}}_{\nu}$ with respect to $\bm{q}$ and $\Omega$, we find that they are in the order of $ \frac{\hbar}{\epsilon_{\textrm{F}}\tau} \ll 1$ and $\hat{\Lambda}^{\textrm{rr}}_\nu$ ($\hat{\Lambda}^{\textrm{aa}}_\nu$) in $\hat{\Pi}^{\textrm{rr}}_\nu$ ($\hat{\Pi}^{\textrm{aa}}_\nu$) can be approximated with $\hat{\sigma}_\nu$ (see Appendices \ref{sec:A}-\ref{sec:A-C} for the detailed calculation). 
Then, by expanding the Fermi distribution function with respect to $\Omega$, the dominant contributions of Eqs. (\ref{eq:3-7}), (\ref{eq:3-8}) and (\ref{eq:3-9}) are written by  
\begin{align}  \label{eq:3-10} 
\hat{\Pi}^{\textrm{ra}}_{\nu} 
	& =  \Omega \sum_{\bm{k}, \omega} f'_{\omega}  \hat{g}^\textrm{r}_{\bm{k}-\frac{\bm{q}}{2},\omega-\frac{\Omega}{2}} \hat{\Lambda}^{\textrm{ra}}_\nu(\omega,\bm{q},\Omega)  \hat{g}^\textrm{a}_{\bm{k}+\frac{\bm{q}}{2},\omega+\frac{\Omega}{2}},
	\\ \label{eq:3-11} 
\hat{\Pi}^{\textrm{rr}}_{ \nu} 
	& = - \sum_{\bm{k}, \omega} \biggl\{ f_{\omega}   \hat{g}^\textrm{r}_{\bm{k}-\frac{\bm{q}}{2},\omega-\frac{\Omega}{2}} \hat{\sigma}_\nu \hat{g}^\textrm{r}_{\bm{k}+\frac{\bm{q}}{2},\omega+\frac{\Omega}{2}} + \frac{1}{2} \Omega f'_{\omega}  \hat{g}^\textrm{r}_{\bm{k}-\frac{\bm{q}}{2},\omega-\frac{\Omega}{2}} \hat{\sigma}_\nu  \hat{g}^\textrm{r}_{\bm{k}+\frac{\bm{q}}{2},\omega+\frac{\Omega}{2}} \biggr\},
	\\ \label{eq:3-11-2} 
\hat{\Pi}^{\textrm{aa}}_{ \nu} 
	& = \sum_{\bm{k}, \omega} \biggl\{ f_{\omega}   \hat{g}^\textrm{a}_{\bm{k}-\frac{\bm{q}}{2},\omega-\frac{\Omega}{2}} \hat{\sigma}_\nu \hat{g}^\textrm{a}_{\bm{k}+\frac{\bm{q}}{2},\omega+\frac{\Omega}{2}} - \frac{1}{2} \Omega f'_{\omega} \hat{g}^\textrm{a}_{\bm{k}-\frac{\bm{q}}{2},\omega-\frac{\Omega}{2}} \hat{\sigma}_\nu  \hat{g}^\textrm{a}_{\bm{k}+\frac{\bm{q}}{2},\omega+\frac{\Omega}{2}}\biggr\}.
\end{align}
In addition, $\hat{\Pi}^{\textrm{rr}}_{ \nu}$ and $\hat{\Pi}^{\textrm{aa}}_{ \nu} $ are shown to be much smaller than $\hat{\Pi}^{\textrm{ra}}_{\nu}$ (The detailed calculation is given in Appendix \ref{sec:D}). 
Thus, $\hat{\Pi}_{\nu} $ is approximately given by
\begin{align} \label{eq:3-12} 
\hat{\Pi}^{\textrm{}}_{\nu} \simeq 
\hat{\Pi}^{\textrm{ra}}_{\nu} 
	& =  \Omega \sum_{\bm{k}, \omega} f'_{\omega}  \hat{g}^\textrm{r}_{\bm{k}-\frac{\bm{q}}{2},\omega-\frac{\Omega}{2}} \hat{\Lambda}^{\textrm{ra}}_\nu (\omega,\bm{q},\Omega)  \hat{g}^\textrm{a}_{\bm{k}+\frac{\bm{q}}{2},\omega+\frac{\Omega}{2}}.
\end{align}
In the low-temperature limit, we approximate the derivative of the Fermi distribution function as $f'_{\omega} = -\delta(\omega)$. 
Then, the integral over $\omega$ in $\hat{\Pi}^{\textrm{ra}}_{\nu} $ reduces to 
\begin{align}  \label{eq:3-13} 
\hat{\Pi}^{\textrm{ra}}_{\nu}(\bm{q},\Omega) 
	& = - \frac{\Omega}{2\pi} \sum_{\bm{k}}  \hat{g}^\textrm{r}_{\bm{k}-\frac{\bm{q}}{2}, -\frac{\Omega}{2}} \hat{\Lambda}^{\textrm{ra}}_\nu(0,\bm{q},\Omega)  \hat{g}^\textrm{a}_{\bm{k}+\frac{\bm{q}}{2},\frac{\Omega}{2}}.
\end{align}
We further expand $\hat{\Lambda}_\nu^{\textrm{ra}}$ as  
$\hat{\Lambda}^{\textrm{ra}}_\nu = \sum_{\alpha=0,x,y,z} \hat{\sigma}_\alpha {\Lambda}^{\textrm{ra}}_{\nu \alpha}$, 
and rewrite Eq. (\ref{eq:3-13}) as
\begin{align}  \label{eq:3-14} 
 \hat{\Pi}^{\textrm{ra}}_{\nu} (\bm{q},\Omega)
	& = -\frac{\Omega}{2\pi} \sum_{\alpha=0,x,y,z} \hat{I}_\alpha(\bm{q},\Omega)  \Lambda^{\textrm{ra}}_{\nu \alpha}(0,\bm{q},\Omega).
\end{align}
Here, we define $\hat{I}_\zeta(\bm{q},\Omega) 
 	 \equiv \sum_{\bm{k}} 
		\hat{g}^\textrm{r}_{\bm{k}-\frac{\bm{q}}{2},-\frac{\Omega}{2}} \hat{\sigma}_\zeta   \hat{g}^\textrm{a}_{\bm{k}+\frac{\bm{q}}{2},\frac{\Omega}{2}} $, which is calculated up to the quadratic terms in $q$ and the primary terms in $\Omega$ as (see Appendices \ref{sec:A} and \ref{sec:B} for the detailed derivation)
\begin{align} \label{eq:3-15} 
\hat{I}_{\zeta = 0} 
	&= \frac{\pi\nu_e}{2\eta} \biggl[ \biggl(1- i\Omega \tau - \frac{1}{2}\ell^2 q^2\biggr)\hat{\sigma}_0 + \sum_{\alpha,u=x,y}\frac{i}{2}\ell \hat{\sigma}_u q_\alpha \epsilon_{u\alpha z} \biggr], 
	       \\  \label{eq:3-16} 
\hat{I}_{\zeta =x,y}
	&= \frac{\pi\nu_e}{2\eta} \biggl[ \sum_{\nu=x,y} \biggl\{ \frac{1}{2}\biggl(1- i\Omega \tau - \frac{3}{4}\ell^2 q^2 \biggr)\delta_{\zeta \nu} + \frac{1}{4}\ell^2 q_\zeta q_\nu  \biggr\} \hat{\sigma}_\nu 
	 + \sum_{\alpha=x,y} \frac{i}{2}\ell \hat{\sigma}_0 q_\alpha \epsilon_{\zeta \alpha z} \biggr],
	 \\  \label{eq:3-16-2} 
\hat{I}_{\zeta=z}
	& = o \left( \frac{\hbar}{\epsilon_\textrm{F} \tau }\right), 
\end{align}
where $\ell=\tilde{v}_{\textrm{F}}\tau$ is the mean free path.
Since $\hat{I}_{\zeta=z}$ is negligibly small as compared with $\hat{I}_{\zeta = 0} $ and $\hat{I}_{\zeta =x,y}$, we consider only the contributions from $\hat{I}_{\zeta=0,x,y}$. 
Since Eqs. (\ref{eq:3-15}) and (\ref{eq:3-16}) do not include $\hat{\sigma}_z$, they are represented by using the Pauli matrices as  
\begin{align} \label{eq:3-17} 
 		\hat{I}_\zeta
	& = \sum_{\mu=0,x,y} I_{\zeta\mu}  \hat{\sigma}_\mu
			+ o\left(\frac{\hbar}{\epsilon_{\textrm{F}}\tau}\right), 
\end{align}
where $I_{\zeta\mu}$ is the 3 $\times$ 3 symmetric matrix given by 
\begin{widetext}
\begin{align} \label{eq:3-18} 
I
	& = \frac{\pi \nu_e}{2\eta}
\left(\begin{tabular}{lcccc}
	& $1 -i\Omega \tau - \frac{1}{2}\ell^2q^2$  & $\frac{i}{2} \ell q_y$  & $-\frac{i}{2} \ell q_x$  \\
	&$\frac{i}{2} \ell q_y$  & $\frac{1}{2} ( 1 - i \Omega \tau  - \frac{1}{2}\ell^2  q^2) + \frac{1}{8}\ell^2(q_x^2-q_y^2)$ & $\frac{1}{4} \ell^2 q_{x}q_{y}$ \\
	&$-\frac{i}{2} \ell q_x$ & $\frac{1}{4} \ell^2 q_{x}q_{y}$ &$\frac{1}{2} ( 1 - i \Omega \tau  - \frac{1}{2}\ell^2  q^2) - \frac{1}{8}\ell^2(q_x^2-q_y^2)$
\end{tabular} \right).
\end{align}
\end{widetext}
On the other hand, from Eq. (\ref{eq:3-4-2}), the vertex function $\hat{\Lambda}^{\textrm{ra}}_\nu$ can be described by
\begin{align}\notag
\hat{\Lambda}^{\textrm{ra}}_\nu 
	& =  \hat{\sigma}_\nu + \sum_{\alpha=0,x,y}\Gamma^{\textrm{ra}}_{\nu \alpha}\hat{\sigma}_\alpha + \sum_{\alpha=0,x,y}[ (\Gamma^{\textrm{ra}})^2]_{\nu \alpha}\hat{\sigma}_\alpha + \cdots 
	\\
	& = \sum_{\alpha=0,x,y}[(1-\Gamma^{\textrm{ra}})^{-1}]_{\nu\alpha} \hat{\sigma}_\alpha,
\end{align}
where the second equality holds when $\max\{ \sum_{\nu} |\Gamma^{\textrm{ra}}_{\mu \nu}| \} <1$ is satisfied. 
By using $1-i\Omega\tau = \frac{1}{1+i\Omega\tau}+ O\left(\Omega^2\right)$, $\Gamma^{\textrm{ra}}=n_i u_i^2I$, and Eq. (\ref{eq:3-18}), one can see $\Gamma^{\textrm{ra}}$ indeed satisfies $\max\{ \sum_{\nu} |\Gamma^{\textrm{ra}}_{\mu \nu}| \} <1$.
Then, the matrix $\Gamma^{\textrm{ra}} \Lambda^{\textrm{ra}} $ is calculated as
\begin{align}\notag
	 \Gamma^{\textrm{ra}}  \Lambda^{\textrm{ra}} 
	& =
	-1+(1-\Gamma^{\textrm{ra}})^{-1}
	\\  \label{eq:3-19} 
	& =
	\left(\begin{array}{ccc}
	 $0$ 
	& $0$   
	& $0$     \\
		 $0$ 				
		& $1$
		& $0$ \\
			 $0$
		 	& $0$
	 		& $1$
	\end{array} \right)
	+
      \left(\begin{tabular}{ccc}
	 $\frac{1 }{q^2 \ell^2 + i\Omega \tau } $
	& $\frac{ i \ell q_y}{q^2 \ell^2 + i\Omega \tau}$  
	& $ - \frac{ i \ell q_x}{q^2 \ell^2 + i\Omega \tau}$    \\
		 $\frac{ i \ell q_y}{q^2 \ell^2 + i\Omega \tau}$ 				
		& $ - \frac{q_y^2 \ell^2 }{q^2 \ell^2 + i\Omega \tau} $
		&  $\frac{ q_x q_y \ell^2 }{q^2 \ell^2 + i\Omega \tau} $ \\
			 $ - \frac{ i \ell q_x}{q^2 \ell^2 + i\Omega \tau}$
	 		& $\frac{ q_x q_y \ell^2 }{q^2 \ell^2 + i\Omega \tau} $
	 		&   $ - \frac{q_x^2 \ell^2 }{q^2 \ell^2 + i\Omega \tau} $
	\end{tabular} \right),
\end{align}
from which we obtain the spin-spin response function as
\begin{align}   \label{eq:3-20} 
 \hat{\Pi}_{\nu} 
	& \simeq -\frac{\Omega\nu_e}{4\eta}  \sum_{\zeta'=0,x,y} [\Gamma^{\textrm{ra}} \Lambda^{\textrm{ra}} ]_{\zeta' \nu} \hat{\sigma}_{\zeta'},
\end{align}
where we have used the fact that $I$ and $\Lambda^{\textrm{ra}}$ are symmetric matrices. 
Thus, from Eqs. (\ref{eq:3-2}) and (\ref{eq:3-20}), the $\mu=x,y$ components of the spin density are given by  
\begin{align}  \label{eq:3-20-2} 
s_\mu 
	& = - \frac{e\tilde{v}_\textrm{F}\nu_e  \tau}{2L^2} \epsilon_{z \nu u } \partial_t\sum_{\bm{q},\Omega} e^{i(\Omega t -\bm{q}\cdot\bm{x})}
	[\Gamma^{\textrm{ra}}\Lambda^{\textrm{ra}}]_{\mu\nu}  A_{{\textrm{em}}, u}.
\end{align}
Substituting Eq. (\ref{eq:3-19}) in Eq. (\ref{eq:3-20-2}), we obtain
\begin{align} \label{eq:3-21} 
s_x 
	& = \frac{1}{2}e\tilde{v}_\textrm{F}\nu_e  \tau E_y 
	 + \frac{e\tilde{v}_\textrm{F}\nu_e  \tau}{2L^2}  \partial_t \sum_{\bm{q},\Omega} e^{i(\Omega t -\bm{q}\cdot\bm{x})}
	\frac{  \ell^2 (q_y^2  A_{{\textrm{em}}, y} + q_y q_x  A_{{\textrm{em}}, x})}{q^2 \ell^2 + i\Omega \tau},
	\\ \label{eq:3-22} 
s_y	& = -\frac{1}{2}e\tilde{v}_\textrm{F}\nu_e  \tau E_x
	 - \frac{e\tilde{v}_\textrm{F}\nu_e  \tau}{2L^2}  \partial_t \sum_{\bm{q},\Omega} e^{i(\Omega t -\bm{q}\cdot\bm{x})}
	\frac{  \ell^2 (q_x^2  A_{{\textrm{em}}, x} + q_y q_x  A_{{\textrm{em}}, y})}{q^2 \ell^2 + i\Omega \tau}.
\end{align}

The second terms of Eqs. (\ref{eq:3-21}) and (\ref{eq:3-22}) can be described by using the charge density $\rho_e$ on the surface.
Here, $\rho_e\equiv e\langle \psi^\dagger \psi \rangle = \frac{i \hbar e^2 \tilde{v}_{\textrm{F}} }{L^2} \epsilon_{z \nu u}\sum_{\bm{q},\Omega}e^{i(\Omega t-\bm{q}\cdot\bm{x})}  {\textrm{tr}}[ \hat{\Pi}^{\textrm{ra}}_{0\nu} ] A_{{\textrm{em}}, u} ( \bm{q},\Omega)$, is obtained from the charge-spin response function $\hat{\Pi}_{0\nu}=\hat{\Pi}_{\nu}$ as 
\begin{align} \notag 
\rho_e 
	&=-
	\frac{i \hbar e^2 \tilde{v}_{\textrm{F}}\nu_e }{2\eta L^2} \epsilon_{z \nu u}\sum_{\bm{q},\Omega}e^{i(\Omega t-\bm{q}\cdot\bm{x})} \Omega 
	\biggl\{ [\Gamma^{\textrm{ra}}\Lambda^{\textrm{ra}}]_{0\nu}  A_{{\textrm{em}}, u} \biggr\}
	\\ 
	&=
	\frac{ e^2 \tilde{v}_{\textrm{F}}\nu_e \tau}{L^2} \ell \partial_t \nabla_\nu \sum_{\bm{q},\Omega}e^{i(\Omega t-\bm{q}\cdot\bm{x})} \frac{ 1 }{q^2 \ell^2 + i\Omega \tau}    A_{{\textrm{em}}, \nu}
	\label{eq:3-23} 
	\\
	&=  - 2 e^2 \nu_e D \tau  \bm{\nabla} \cdot \langle \bm{E} \rangle_{\textrm{D}}, 
	\label{eq:3-24} 
\end{align}
where $ D \equiv \frac{1}{2} \tilde{v}_{\textrm{F}}^2 \tau = \frac{1}{2} \tilde{v}_{\textrm{F}} \ell$ is the diffusion constant. 
Here, $\langle \bm{E} \rangle_{\textrm{D}}$ is defined by the convolution of $\bm{E}$  and the diffusive propagation function $\mathcal{D}$ as 
\begin{align} \label{eq:3-25} 
\langle \bm{E}  \rangle_{\textrm{D}}
	& \equiv \frac{1}{\tau}\int_{-\infty}^\infty dt' \int d\bm{x'}  \mathcal{D} (\bm{x}-\bm{x'},t-t') \bm{E} (\bm{x'}, t'), 
	\\ \label{eq:3-26} 
\mathcal{D} (\bm{x},t)
	& = \frac{1}{L^2} \sum_{\bm{q},\Omega}e^{i(\Omega t-\bm{q}\cdot\bm{x})}  \frac{1}{ 2 D q^2  + i\Omega } 
	\\ \label{eq:3-27} 
	&\sim \frac{\theta(t)}{8\pi Dt}\exp\left[-\frac{|\bm{x}|^2}{8Dt}\right].
\end{align}
The diffusive propagation function $\mathcal{D}$ is also the Green's function satisfying the following differential equation 
\begin{align} \label{eq:3-28} 
\left( \partial_t  - 2 D \nabla^2 \right)  \mathcal{D} (\bm{x},t) & = \delta(\bm{x}) \delta(t).
\end{align}
Equations (\ref{eq:3-24}) and (\ref{eq:3-25}) show that due to the impurities the effect of the applied electric field on the surface electrons is not instantaneous but diffusively propagates on the surface of the TI. 
Equation (\ref{eq:3-25}) gives the definition of such a nonlocal electric field.
Suppose that the surface of the TI is isotropic, the gradient of the charge density is given by 
\begin{align}   \label{eq:3-29} 
 \nabla_x \rho_e
	& = 	-\frac{ e^2 \tilde{v}_{\textrm{F}}\nu_e \tau}{L^2} \partial_t \sum_{\bm{q},\Omega}e^{i(\Omega t-\bm{q}\cdot\bm{x})}
	 \frac{  \ell (q_x^2  A_{{\textrm{em}}, x} + q_y q_x  A_{{\textrm{em}}, y})}{q^2 \ell^2 + i\Omega \tau}, 
	 \\   \label{eq:3-30} 
 \nabla_y \rho_e
	& = 	- \frac{e^2 \tilde{v}_{\textrm{F}}\nu_e \tau}{L^2} \partial_t \sum_{\bm{q},\Omega}e^{i(\Omega t-\bm{q}\cdot\bm{x})}
	 \frac{  \ell (q_y^2  A_{{\textrm{em}}, y} + q_y q_x  A_{{\textrm{em}}, x})}{q^2 \ell^2 + i\Omega \tau} ,
\end{align}
which is related to the spin density [Eqs. (\ref{eq:3-21}) and (\ref{eq:3-22})] as 
\begin{align}\label{eq:3-31} 
\bm{s} &=  \frac{1}{2}e\tilde{v}_\textrm{F}\nu_e  \tau (\bm{E}\times \bm{z}) + \frac{\ell}{2e} \left( \bm{z} \times \bm{\nabla} \right)  \rho_e. 
\end{align}
On the other hand, the spin density on the surface of the TI is related to the charge current via
\begin{align}\label{eq:3-33} 
j_\mu=-\frac{\partial \mathcal{H}_{\rm em}}{\partial A_{{\rm em},\mu}}=2e\tilde{v}_{\rm F}(\bm{z}\times {\bm s})_\mu,
\end{align}
where we have replaced $v_{\rm{F}}$ in Eq. (\ref{eq:2-3}) with $\tilde{v}_{\rm{F}}$ so as to take into account the effects of impurities.
By substituting Eq. (\ref{eq:3-31}) in Eq. (\ref{eq:3-33}), we obtain 
\begin{align} \label{eq:3-32} 
\bm{j} 
	& = e^2 \tilde{v}_\textrm{F}^2 \nu_e \tau \bm{E} 
		-  \tilde{v}_\textrm{F} \ell \bm{\nabla} \rho_e.
\end{align}
We have confirmed that Eqs. (\ref{eq:3-24}) and (\ref{eq:3-32}) satisfy the charge conservation law: $\dot{\rho}_e + \bm{\nabla}\cdot \bm{j} =0$ (see Appendix \ref{sec:F} for the detailed calculation).

\section{Properties of the charge, spin, charge current, and spin current densities} \label{sec:4} 
Using the results in Sec. \ref{sec:3}, 
we discuss the property of the charge, spin, charge current, and spin current densities induced by the electric field applied on the disordered surface of the doped TI.

\subsection{Charge density}
We find that as shown in Eq. (\ref{eq:3-24}) the charge density $\rho_e$ is induced by the divergence of the nonlocal electric field:$\langle \bm{\nabla}\cdot \bm{E} \rangle_\textrm{D}$.
Therefore, when we apply a uniform electric field, no charge density is induced.
From Eqs. (\ref{eq:3-25})-(\ref{eq:3-27}) we obtain the diffusion equation for the charge transport:
\begin{align} \label{eq:3-3-1} 
\left( \partial_t  - 2 D \nabla^2 \right) \rho_e (\bm{x},t) 
	& =   -2 e^2 \nu_e D   \bm{\nabla} \cdot \bm{E}(\bm{x},t),
\end{align}
which indicates that the divergence of the applied electric field works as a source of the diffusive propagation of the charge density. 
We find that from the left side of the equation above, $\left( \partial_t  - 2 D \nabla^2 \right) \rho_e$, the diffusion constant is $2D$, a twice of that on the surface of a metal\cite{rf:sakai14}. 
The factor $2$ comes from the difference in the self-energy due to impurity scattering: 
The self-energy on the surface of an isotropic metal is given by $\pi n_{\textrm{i}} u_{\textrm{i}}^2  \nu_{e,m} $, 
where $\nu_{e,m}$ is the density of states in the metal, 
whereas that on the surface of the TI is $\frac{1}{2}\pi n_{\textrm{i}} u_{\textrm{i}}^2  \nu_e $; 
The factor $\frac{1}{2}$, which originates from the linear dispersion of the surface of the TI, leads to the coefficient $2D$. 
Here, the diffusive equation of motion qualitatively agrees with the previous works\cite{rf:Burkov10,rf:Schwab11,rf:Taguchi14}.

\subsection{Spin density}\label{sec:4-2}
We turn to the discussion on the spin density given by Eq. (\ref{eq:3-31}).
We find that the spin density can be divided into the one induced by the local electric field $\bm{E}$ and that by the nonlocal electric field $\langle \bm{E} \rangle_{\textrm{D}}$. 
We define the local spin density $\bm{s}^{(\textrm{l})}$ as the first term of Eq. (\ref{eq:3-31}):  
\begin{align} \label{eq:local-spin}
\bm{s}^{(\textrm{l})} = \frac{1}{2}e\tilde{v}_\textrm{F}\nu_e  \tau (\bm{E}\times \bm{z}).
\end{align}
The local spin density lies in the $xy$ plane and is perpendicular to the local electric field $\bm{E}$\cite{rf:Raghu10,rf:Misawa11,rf:Liu13}, 
which is the consequence of the spin-momentum locking on the surface of the TI.  
The spin due to the local electric field is a kind of the Edelstein effect\cite{rf:Edelstein90}.
%
%
On the other hand, 
the nonlocal spin density $\bm{s}^{(\textrm{nl})} =  \frac{\ell}{2e} ( \bm{z} \times  \bm{\nabla})  \rho_e$, the second term of Eq. (\ref{eq:3-31}), 
is generated from the spatial gradient of the charge density.
The direction of $\bm{s}^{(\textrm{nl})}$ is in the $xy$ plane and perpendicular to the gradient of the charge density $\bm{\nabla}\rho_e$. 
Here, $\bm{s}^{(\textrm{nl})}$ is also written in terms of $\langle \bm{E}\rangle_\textrm{D}$ as 
\begin{align} 
\label{eq:spin-x10} 
\bm{s}^{(\textrm{nl})} 
	& =  -\frac{e\tilde{v}_\textrm{F} \nu_e\tau \ell^2 }{2} \left( \bm{z} \times \bm{\nabla} \right)(\bm{\nabla} \cdot \langle \bm{E} \rangle_{\textrm{D}}),
\end{align}
which means that the spin density is generated by the second spatial derivative of the nonlocal electric field $\langle \bm{E}\rangle_\textrm{D}$. 
%

The nonlocal spin density diffusively propagates through the impurity scatterings on the surface of the TI. 
The diffusion propagation of the spin is described by 
\begin{align} \label{eq:spin-diffusion-propagation}
 (\partial_t - 2D \nabla^2  ) \bm{s}^{(\textrm{nl})}
	& =  -\frac{e\tilde{v}_\textrm{F} \nu_e \ell^2 }{2} \left( \bm{z} \times \bm{\nabla} \right)(\bm{\nabla} \cdot \bm{E}). 
\end{align}
We find that the diffusion propagation of the spin is triggered by an inhomogeneous electric field, $\bm{\nabla}(\bm{\nabla} \cdot \bm{E})$. 
(This property is also predicted in Ref. $34$). 
Hence, when we apply a uniform electric field on the surface, the nonlocal spin density is not generated.
As in the case of charge density, the diffusion constant for the spin density is $2D$.

We note that both the local and nonlocal spin densities are proportional to $\tilde{v}_{\textrm{F}}$.
Since $\tilde{v}_{\textrm{F}}$'s on the top and the bottom sides of the TI have opposite signs, the direction of the induced spin on the top surface of the TI is perfectly opposite to that on the bottom surface of the TI, 
when we apply the same electric field on both the top and bottom side of the TI.

\subsection{Charge current}
The charge current on the surface of the TI is proportional to the spin density as shown in Eq. (\ref{eq:3-33})\cite{rf:Ando13}.
The origin lies on the spin-momentum locking on the surface of the TI. 
From Eq. (\ref{eq:3-32}), we find that the charge current $\bm{j}$ can also be divided into the local and nonlocal parts. 
The first term of Eq. (\ref{eq:3-32}) gives the local charge current $\bm{j}^{(\textrm{l})} =  e^2 \tilde{v}_\textrm{F}^2 \nu_e \tau \bm{E}$.
$\bm{j}^{(\textrm{l})}$ is the electric current directly induced by the applied electric field.
We find from the above result that the longitudinal conductivity is given by $j_\mu/E_\mu =  e^2 \tilde{v}_\textrm{F}^2 \nu_e \tau$, 
which agrees with the existing work\cite{rf:Misawa11}.   
The second term of Eq. (\ref{eq:3-32}) corresponds to the nonlocal charge current density $\bm{j}^{(\textrm{nl})}=- \tilde{v}_\textrm{F} \ell \bm{\nabla} \rho_e$.
$\bm{j}^{(\textrm{nl})}$ is the diffusion current and is generated by the spatial gradient of the charge density on the disordered surface of the doped TI. 
The charge current can be rewritten by using the nonlocal electric field $\langle \bm{E} \rangle_\textrm{D}$ as 
\begin{align}  \label{eq:4-3-2} 
\bm{j}^{(\textrm{nl})}=  e^2 \tilde{v}^2_\textrm{F} \ell^2 \nu_e \tau \bm{\nabla}  (\bm{\nabla}\cdot \langle \bm{E} \rangle_\textrm{D}).
\end{align}
Since both the local and nonlocal charge currents are proportional to $\tilde{v}_{\textrm{F}}^2$, their directions are the same for top and bottom surface of the TI.

\subsection{Spin current}
Next, we calculate the spin current due to the applied electric field on the disordered surface of the doped TI.
The spin current $j_{i}^\alpha$ is defined by 
\begin{align}  \label{eq:4-4-1} 
\dot{s}^\alpha & + \nabla_i j_{i}^\alpha  =  \mathcal{T}^\alpha,
\end{align}
where the subscript and superscript of $j_{i}^\alpha$ denote the direction of the flow and spin, respectively, 
and $\mathcal{T}^\alpha$ represents the spin relaxation. 
Using the Hamiltonian in Eq. (\ref{eq:2-1}) and the Heisenberg equation for $s^\alpha$, we obtain 
\begin{align}  \label{eq:4-4-2} 
j_{i}^\alpha  
	= & \frac{\tilde{v}_{\textrm{F}}}{2e} \epsilon_{z \alpha i } \rho_e.
\end{align}
Note that the direction of the flow and spin is perpendicular each to other. 
This is the consequence of the spin-momentum locking on the surface of the TI.
We also note that the spin current is proportional to the charge density, and from Eq. (\ref{eq:3-24}), proportional to the divergence of the nonlocal electric field: 
\begin{align}  \label{eq:4-4-3} 
j_{i}^\alpha  
		= &  - e\tilde{v}_{\textrm{F}} \nu_e D \tau \epsilon_{z \alpha i}   \bm{\nabla} \cdot \langle \bm{E} \rangle_{\textrm{D}} .
\end{align}
Hence, when we apply a spatially uniform electric field on the surface, no spin current is induced.
Besides, we find that the spin current is an odd-function of $\tilde{v}_{\textrm{F}}$, which means the spin current depends on the chirality on the surface of the TI. 
Namely, the relative direction between flow and spin of $j_{i}^\alpha$ on the top side of the TI is opposite to that on the bottom side of the TI. 

In the conventional spin-orbit coupled systems, the spin current is generated by an applied electric field\cite{rf:Murakami03,rf:Sinova04,rf:Rashuba03,rf:Kimura04,rf:Kato04}, which called the spin Hall effect. 
Besides, the generated spin current can be converted into the charge current via the spin-orbit interaction.\cite{rf:Saitoh06}  
These effect can be understand from the coupling between the spin current and the charge current: $j_i\propto \epsilon_{ij\alpha}j_j^\alpha$\cite{rf:Saitoh06}.
On the surface of the TI, on the other hand, 
we find from Eqs. (\ref{eq:4-3-2}) and (\ref{eq:4-4-3}) that the nonlocal charge current is proportional to the gradient of the spin current\cite{rf:Taguchi15}:
\begin{align}  \label{eq:4-4-4} 
\bm{j}^{(\textrm{nl})}  
		= &  -e\ell \epsilon_{z\alpha i} \bm{\nabla} j_{i}^\alpha.
\end{align}
Again, this is the consequence of the spin-momentum locking and Eq. (\ref{eq:4-4-4}) generally holds for the system of electrons on a surface of TIs\cite{rf:Taguchi15}. This property in the TI is distinct from that in a conventional metal.
The direction of the charge current is parallel to the spatial gradient of $j_{i}^\alpha$.
This relation is plausible due to the following reasons.
First, the charge density $\rho_e$ is proportional to the spin current.
Second, a diffusive particle current generally proportional to a spatial gradient of particles.  
We note that there is no relation between the spin current and the local charge current $\bm{j}^{(\textrm{l})}$.

Finally, we comment on the property of the spin relaxation torque.
The relaxation torque $\mathcal{T}^\alpha$ defined in Eq. (\ref{eq:4-4-1}) is obtained within the linear response to the electric field as 
\begin{align}  \notag 
\mathcal{T}^\alpha 
	 =& \frac{1}{2}e\tilde{v}_\textrm{F}\nu_e  \tau (\dot{\bm{E}}\times \bm{z})_\alpha  
	 	- \left(\frac{\ell}{2e} \partial_t + \frac{\tilde{v}_\textrm{F}}{2e} \right) \left( \bm{z} \times \bm{\nabla}\right)_\alpha  \rho_e
	\\   \label{eq:4-4-5} 
	 =&
	  \frac{1}{2}e\tilde{v}_\textrm{F}\nu_e \left[ \tau (\dot{\bm{E}}\times \bm{z})_\alpha  
	 	 + 2D\tau  \left( \bm{z} \times \bm{\nabla} \right)_\alpha  (  \bm{\nabla} \cdot \langle \bm{E} \rangle_{\textrm{D}} ) \right]
		  +o[ (\bm{z} \times \bm{\nabla})_\alpha (\bm{\nabla} \cdot \langle \dot{\bm{E}} \rangle_{\textrm{D}}) ].
\end{align}
Here, $\mathcal{T}^\alpha $ can be divided into the local and nonlocal terms, which correspond to the first and second terms, respectively, in the first square bracket in the most right-hand side of Eq. (\ref{eq:4-4-5}).
The local one is given by the time derivative of the applied electric field, and its direction is perpendicular to both $\dot{\bm{E}}$ and $\bm{z}$.
The nonlocal one is proportional to the second derivative of the nonlocal electric field $\langle \bm{E} \rangle_\textrm{D} $.
These above results and properties are the same as the spin density on the surface of the TI with magnetism\cite{rf:Taguchi15}.

\section{Responses to the electric field of a twisted light beam}\label{sec:5} 
Using the results obtained in Sec. \ref{sec:4}, 
we discuss the properties of the spin and charge densities due to the electric field of a twisted light beam with various orbital angular momentum.

\subsection{Electric field of a twisted light beam} \label{sec:5-1} 
\begin{figure}[htbp]\centering 
\includegraphics[scale=0.35]{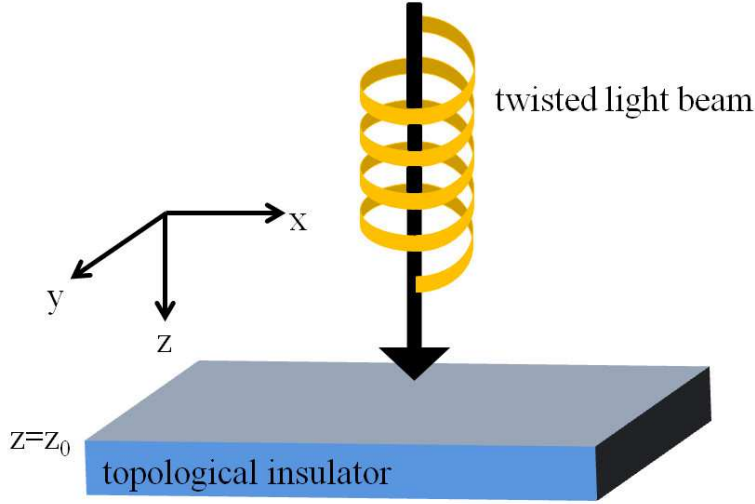}
\caption{(color online) Schematic illustration of the system. The optical twisted light beam is applied to the surface of the topological insulator for normal incidence.
}\label{fig:fig1} 
\end{figure}

First, we explain the property of the electric field of the twisted light beam with the Laguerre-Gaussian modes\cite{rf:Allen,rf:Padgett04}.
%
We assume that the twisted light beam propagates along the $z$ axis and 
the electric field of the twisted light beam lies in the $xy$ plane at the top surface of the TI ($z=z_0$). 
The schematic of the system is illustrated in Fig. \ref{fig:fig1}.
The twisted light beam satisfies the wave equation, $\nabla^2 \bm{E} - \frac{1}{c^2_0} \frac{\partial ^2 \bm{E}}{\partial t^2}=0$, where $c_0$ is the velocity of light in a vacuum. 
Then, the electric field $\bm{E}(\bm{x},t)=(E_x(r,\varphi,t), E_y(r,\varphi,t))$ on the surface is written by\cite{rf:Allen,rf:Padgett04}
\begin{align}
	\label{eq:EF-of-opticalvotex}
\bm{E} & = \mathcal{E} 
		\bigl(       \cos{( \Theta_\mathcal{R} + m^z_{\textrm{L}} \varphi - \Omega t)}  
			 , -\sigma^z_{\textrm{L}} \sin{(\Theta_\mathcal{R}  + m^z_{\textrm{L}} \varphi - \Omega t)} 
		\bigr),
\end{align}
where $r=\sqrt{x^2 +y^2}$ is the distance from the center of the light on the top surface ($z=z_0$) and 
$\varphi= \arctan{(y/x)}$ is the azimuthal angle. 
The helicity $\sigma^z_{\textrm{L}}=+1 (-1)$ denotes the right-hand (left-hand) circularly polarized light and corresponds to the spin angular momentum of light $+1 (-1)$.
The orbital angular momentum of light, $m^z_{\textrm{L}} =0, \pm1, \cdots$, determines the whirling pattern of the electric field on the plane at $z=z_0$, 
which can be manipulated in experiments\cite{rf:Padgett04}.  
%
The phase $\Theta_\mathcal{R} \equiv \Theta_\mathcal{R}(r, z_0) $ depends on $r$ 
and 
the distance from the light source, $z$, 
and is given by\cite{rf:Lembessis09} 
\begin{align} \label{eq:5-1}
\Theta_\mathcal{R} & = -(1+2p+|m^z_{\textrm{L}}|)\tan^{-1}{\biggl[ \frac{z_0}{z_r} \biggr]}  - \frac{q_z r^2 }{2R(z_0)},
\end{align}
where the first term denotes Guoy phase, $R(z)\equiv  z [ 1+ ( z_r /z )^2 ]$ is the radius of the beam curvature with taking the origin of the $z$ axis at the beam waist,  
$z_r = \pi d_0^2/\lambda$ is the Rayleigh range, 
$d_0$ is the waist size of the $m^z_{\textrm{L}}=p=0$ mode,
$\lambda$ is the wavelength,
and 
$q_z$ is the wave vector of the light. 
The integer $p$ is one of the indices that specify the Laguerre-Gaussian mode and denotes the number of oscillations of the electric field $\mathcal{E}$ in the radial direction. 
Here, $\mathcal{E} $ is given by 
\begin{align} 	\label{eq:mathcal(E)}
\mathcal{E} (r, z_0)
	& = \mathcal{E}_0 \sqrt{\frac{2p!}{\pi (p + |m^z_{\textrm{L}}|)![1 + (z_0/z_r)^2]}} (\sqrt{2}u)^{|m^z_{\textrm{L}}|} L_{|m^z_{\textrm{L}}|}^{p} (2u^2) \exp{ (- u^2)},
\end{align}
where $u= r / [ d_0 [1 + (z_0/z_r)^2]^{\frac{1}{2} }]$, $\mathcal{E}_0$ is a constant, and $L_{|m^z_{\textrm{L}}|}^{p}(y)$ is the Laguerre polynomials defined by 
\begin{align}
L_{|m^z_{\textrm{L}}|}^p (y) & = \sum_{k=0}^p (-1)^k \frac{(|m^z_{\textrm{L}}|+p)!}{(p-k)! (k+|m^z_{\textrm{L}}|)! k! }y^k .
\label{eq:Laguerre}
\end{align}
In the following discussion, we consider only the $p=0$ modes.
We also assume that the twisted light beam is focused at the surface of the TI, i.e., $z_0=0$. Then, the phase $\Theta_\mathcal{R}$ becomes zero, and the magnitude $\mathcal{E}$ defined in Eq. (\ref{eq:mathcal(E)}) reduces to
\begin{align} 	\label{eq:mathcal(E)p=0}
\mathcal{E} (r, 0)
	& = \mathcal{E}_0 \sqrt{\frac{2}{\pi |m^z_{\textrm{L}}|!}} \left(\frac{\sqrt{2}r}{d_0}\right)^{|m^z_{\textrm{L}}|} \exp{ \left(-\frac{r^2}{d_0^2}\right)}.
\end{align}

\begin{figure}[htbp]\centering 
\includegraphics[scale=0.4]{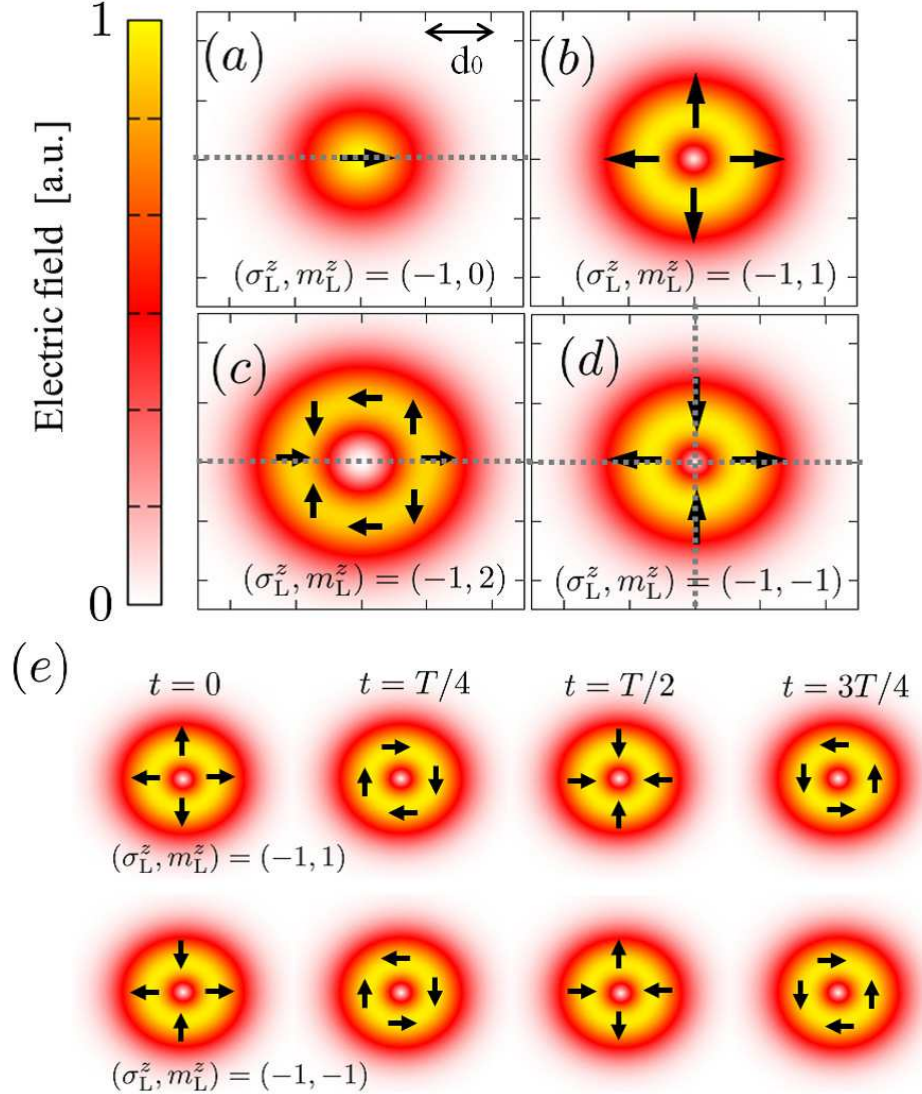}
\caption{(color online) (a)-(d) Snapshots of the electric field induced by the twisted light beam with $(a) (\sigma^z_{\textrm{L}},m^z_{\textrm{L}})=(-1,0)$, $(b) (-1,1)$, $(c) (-1,2)$, and $(d)  (-1,-1)$, where $\sigma_{\textrm{L}}^z$ and $m_{\textrm{L}}^z$ denote the spin and orbital angular momentum of light, respectively. Shown are density plots of the magnitude of the electric field, and the black arrows show the direction of the electric field. 
$d_0$ is the waist size for the $m^z_{\textrm{L}}=0$ mode.
(e) Time evolution of the electric field with $(\sigma_{\textrm{L}}^z,m_{\textrm{L}}^z)=(-1,1)$ (top) and $(-1,-1)$ (bottom), where $T=2\pi/\Omega$ with $\Omega$ being the angular frequency of the light. 
}\label{fig:fig2} 
\end{figure}

Figures \ref{fig:fig2}(a)-\ref{fig:fig2}(d) show the snapshots of the electric field for $\sigma_{\textrm{L}}^z=-1$ and $m_{\textrm{L}}^z=0, 1, 2,$ and $-1$.
In both cases of $m_{\textrm{L}}^z=0$ and $m_{\textrm{L}}^z\neq 0$, the amplitude of the electric field exponentially decays with $r^2$. In addition to this, for the cases of nonzero $m_{\textrm{L}}^z$, 
the magnitude of the electric field vanishes at $r=0$ because of the phase singularity.
This is a characteristic property of the twisted light beam.
Besides, the direction of the electric field depends on the polar angle around the center of the incident light:
While the direction of the electric field is uniform for $(\sigma_{\textrm{L}}^z,m_{\textrm{L}}^z)=(-1,0)$ [Fig. \ref{fig:fig2}(a)], the direction of the electric field at $(\sigma_\textrm{L}^z, m_\textrm{L}^z)=(-1,1)$ rotates by $2\pi$ in the counter-clockwise direction as one goes around the beam center from $\varphi=0$ to $2\pi$ [Fig. \ref{fig:fig2}(b)].
Similarly for the cases of $(\sigma_{\textrm{L}}^z,m_{\textrm{L}}^z)=(-1,2)$ and $(-1,-1)$, the direction of the electric field changes by $4\pi$ and $-2\pi$, respectively [Figs. \ref{fig:fig2}(c) and \ref{fig:fig2}(d)]. 
Note that the configurations shown in Figs. 3(a)--3(d) are snapshots and they evolve in time depending on $\sigma_{\textrm{L}}^z$: $\sigma_{\textrm{L}}^z=-1$ means that the electric field at a fixed point rotates in the clockwise direction as time evolves [Fig. \ref{fig:fig2}(e)].

The topological properties of the twisted light beam discussed above can be understood by introducing the winding number of the electric field. In general,  the winding number of a 2D vector field $\bm n=(n_x,n_y)$ on a closed loop $C$ is defined by
\begin{align}\label{eq:winding-number}
\omega_v[\bm{n}] \equiv \frac{1}{2\pi}\oint_{C} \delta_{ij}\epsilon^{\mu\nu}dx_i \frac{n_\mu}{|\bm{n}|} \frac{\partial}{\partial x_j} \left(\frac{n_\nu}{|\bm{n}|}\right),
\end{align}
where $\epsilon^{ij}$ is the 2D Levi-Civita symbol, and $|\bm n|$ is supposed to be nonzero on $C$. The winding number corresponds to the number of times the 2D unit vector $\bm n/|\bm n|$ rotates about the $z$ axis as one traces the contour $C$. For the electric field given by Eq. (\ref{eq:EF-of-opticalvotex}), the winding number $w_v({\bm E})$ is defined on a contour that encloses $r=0$. Substituting Eq. (\ref{eq:EF-of-opticalvotex}) in Eq. (\ref{eq:winding-number}), we obtain $w_v({\bm{E}})=-\sigma_{\textrm{L}}^z m_{\textrm{L}}^z$. 
For example, for the electric field with $(\sigma_{\textrm{L}}^z,m_{\textrm{L}}^z)=(-1,0), (-1,1), (-1,2)$, and $(-1,-1)$, we have $w_v(\bm{E})=0, 1, 2$, and $-1$, respectively, which are consistent with the configurations shown in Fig. \ref{fig:fig2}. 
The result $w_v(\bm{E})=-\sigma_{\textrm{L}}^z m_{\textrm{L}}^z$ is also consistent with the fact that the direction of the electric field with $\sigma_{\textrm{L}}^z=0$ is spatially uniform even for $m_{\textrm{L}}^z\neq 0$ and that the whirling direction for $\sigma_{\textrm{L}}^z=1$ is opposite to that for $\sigma_{\textrm{L}}^z=-1$.

%


\subsection{Charge density}\label{sec:5-2} 
We consider the charge density due to the electric field of the twisted light beam on the disordered surface of the TI.
The setup we consider is schematically described in Fig. \ref{fig:fig1}.
The induced charge density $\rho_e$ is given by Eq. (\ref{eq:3-24}), which can be rewritten as
\begin{align}
\rho_e (\bm{x}, t)
	& = \frac{1}{\tau}\int_{-\infty}^\infty dt' \int d\bm{x'}  \mathcal{D} (\bm{x'}, t') \bar{\rho}_e (\bm{x}-\bm{x'},t-t'),
		\label{eq:rho_e_Laguerre}
\\
\bar{\rho}_e(\bm{x}, t)
	& = -2 e^2\nu_e D \tau \bm{\nabla}\cdot \bm{E}  (\bm{x}, t).
	\label{eq:rho_e'}
\end{align}
By using Eq. (\ref{eq:EF-of-opticalvotex}), 
$\bm{\nabla} r = \bm{r}/r = (\cos{\varphi}, \sin{\varphi}) $
and 
$ \bm{\nabla} \varphi = (\bm{z} \times \bm{r}) /r^2 = ( -\sin{\varphi}/r, \cos{\varphi}/r)$ the divergence of the electric field for $\sigma^z_{\textrm{L}}=\pm1$ is given by
\begin{align}
\bm{\nabla}\cdot \bm{E} 
	& =  \biggl( \frac{\partial \mathcal{E} }{\partial r} - \frac{\sigma^z_{\textrm{L}}m^z_{\textrm{L}} \mathcal{E}}{r}  \biggr)  \cos{[ (m^z_{\textrm{L}}+ \sigma^z_{\textrm{L}})\varphi -\Omega t]}.
	\label{eq:nabla-E}
\end{align}
Then, $\bar{\rho}_e$ becomes 
\begin{align}\label{eq:rho_e'2}
\bar{\rho}_e(\bm{x}, t)
	& =  -2e^2\nu_e D \tau \left( \frac{\partial \mathcal{E} }{\partial r} - \frac{\sigma^z_{\textrm{L}}m^z_{\textrm{L}} \mathcal{E}}{r} \right) \cos{(j^z_{\textrm{L}}\varphi -\Omega t)}, 
\end{align}
where $j^z_{\textrm{L}} \equiv m^z_{\textrm{L}}+ \sigma^z_{\textrm{L}}$ denotes the total angular momentum of light. 

We further simplify Eqs. (\ref{eq:rho_e_Laguerre}) and (\ref{eq:rho_e'}). 
First of all, these equations are valid only for $\Omega\tau\ll1$. This condition means that the period of the oscillation of $\bar{\rho}_e$, which is the same as that of the electric field, $T=2\pi/\Omega$, is much slower than the electron relaxation time $\tau$. On the other hand, the length scale of the spatial variation of $\bar{\rho}_e$ is in the order of the beam waist $d_0$ [see Eq. (\ref{eq:mathcal(E)}) and Fig. \ref{fig:fig2}], which is comparable to the wavelength $\lambda$ of the light. Since $\lambda$ satisfies $\ell/\lambda=2\pi\ell\Omega/c_0=(2\pi\tilde{v}_{\textrm{F}}/c_0)\Omega\tau\ll 1$, the spatial variation of $\bar{\rho}_e$ is much slower than the mean-free path $\ell$, where $c_0$ is the speed of light and we have used $2\pi\tilde{v}_{\rm F}\ll c_0$ for realistic TIs\cite{rf:Taskin12a}. 
Then, since $\tau$ and $\ell$ determines the decay time and decay length of the diffusion propagator $\mathcal{D}(\bm{x},t)$, respectively, $\bar{\rho}_e(\bm x-\bm x', t-t')$ in the integrand of Eq. (\ref{eq:rho_e_Laguerre}) can be approximated as $\bar{\rho}_e(\bm{x},t)$, and the convolution can be approximately described by 
\begin{align} \label{eq:5-2-3} 
\rho_e (\bm{x}, t)
	& \simeq  \alpha \bar{\rho}_e (\bm{x},t), 
\end{align}
where $\alpha \equiv \frac{1}{\tau}\iint dt  d\bm{x}  \mathcal{D} (\bm{x}, t) $ is a constant coefficient and is estimated by 
\begin{align}
\alpha \simeq \frac{1}{\tau}\int_0^{\tau} dt  \int_0^{2\pi} d\phi \int_{0}^{\ell} r dr \mathcal{D}.
\end{align}
%
\begin{figure}[htbp]\centering 
\includegraphics[scale=0.4]{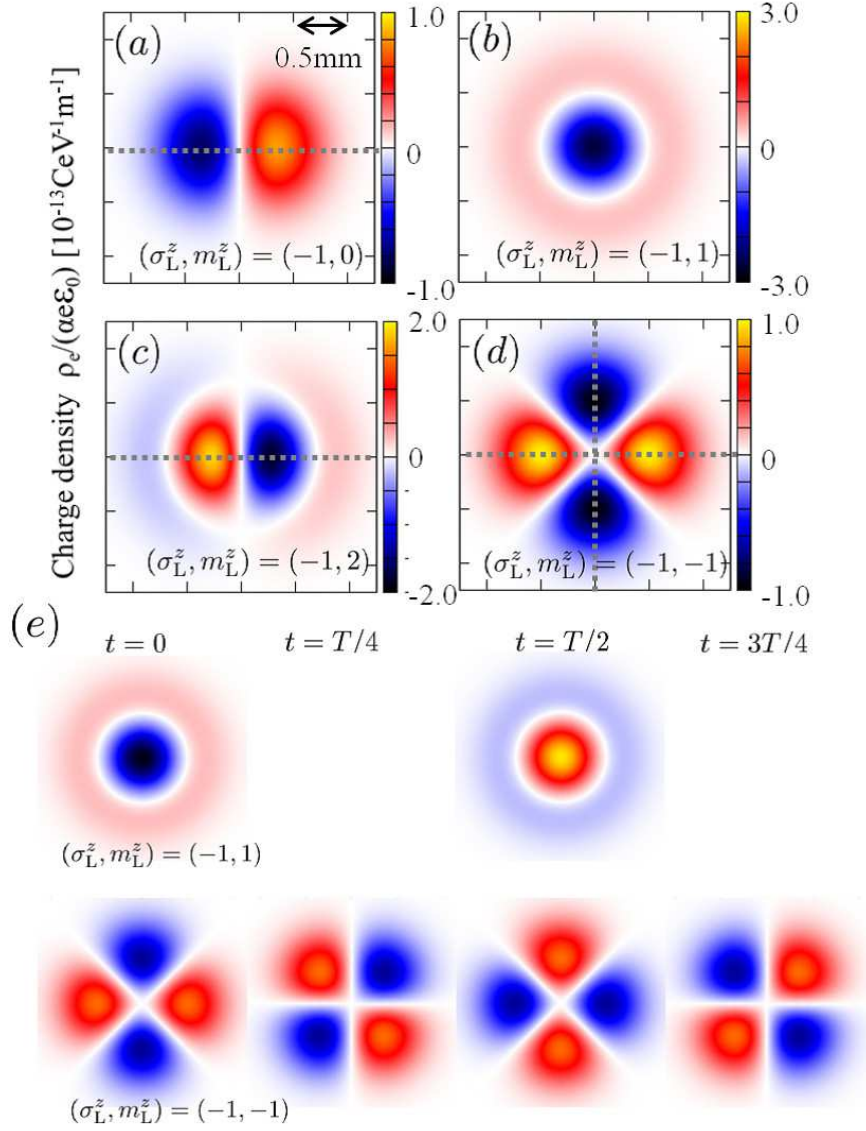}
\caption{(color online) Snapshots of the charge density induced by the optical twisted light beam with $(a) (\sigma^z_{\textrm{L}}, m^z_{\textrm{L}})=(-1,0)$, $(b)(-1,1)$, $(c) (-1,2)$, and $(d) (-1,-1)$.
The dashed lines in (a), (b), and (c) indicate axes of the inversion symmetry. 
(e) Time evolution of the charge density with $(\sigma_{\textrm{L}}^z,m_{\textrm{L}}^z)=(-1,1)$ (top) and $(-1,-1)$ (bottom). 
In all figures, we use $d_0=0.5$ mm, $\epsilon_{\textrm{F}}=100$ meV, $\tilde{v}_{\textrm{F}}=3\times 10^5$ m/s, and $\tau=1\times 10^{-13}$ s.
}\label{fig:fig3} 
\end{figure}
%

Figure \ref{fig:fig3} represents the nonlocal charge density due to the twisted light beam, $\rho_e$, for a $\sigma^z_{\textrm{L}}=-1$ and  $m^z_{\textrm{L}}=0, 1, 2,$ and $-1$.
We find that the distribution of $\rho_e$ depends on the $z$-component of the total angular momentum of light,  $j^z_{\textrm{L}}$.
For $j^z_{\textrm{L}}=0$ [Fig. \ref{fig:fig3}(b)], the charge density is isotropically induced from the center and the sign of the induced charge changes at $r\sim d_0$.
On the other hand, 
the charge density for nonzero $j_{\textrm{L}}^z$ distributes anisotropically. 
The symmetry of the distribution of $\rho_e$ with $|j_{\textrm{L}}^z|=1$ [Figs. \ref{fig:fig3}(a) and \ref{fig:fig3}(c)] and $|j_{\textrm{L}}^z|=2$ [Fig. \ref{fig:fig3}(d)] are the same as that for the electron wave functions with the $p_x$ and $d_{x^2-y^2}$ orbital, respectively.  
The dashed lines in Figs. \ref{fig:fig3}(a), \ref{fig:fig3}(c) and \ref{fig:fig3}(d) indicate the axes of the inversion symmetry of the charge density.
As time evolves, the distribution of the induced charge density rotates around the beam center ($j_{\textrm{L}}^z\neq 0$) or oscillates around the zero value $(j_{\textrm{L}}^z=0)$ with the frequency of light $\Omega$ [see Fig. \ref{fig:fig3}(e)]. 
The time evolution of the charge density also depends on the total angular momentum of light: 
When the sign of the total angular momentum is minus (plus), the distribution of the charge density rotates in the clockwise (counterclockwise) direction around the phase singularity during the irradiation.
When we turn off the incident light, 
the charge density diffusively propagates on the disordered surface of the TI with obeying the diffusive equation of motion represented in Eq. (\ref{eq:3-3-1}). 
Eventually, the induced charge vanishes.

\subsection{Spin density}\label{sec:5-3} 
\begin{figure*}[htbp]\centering 
\includegraphics[scale=0.3]{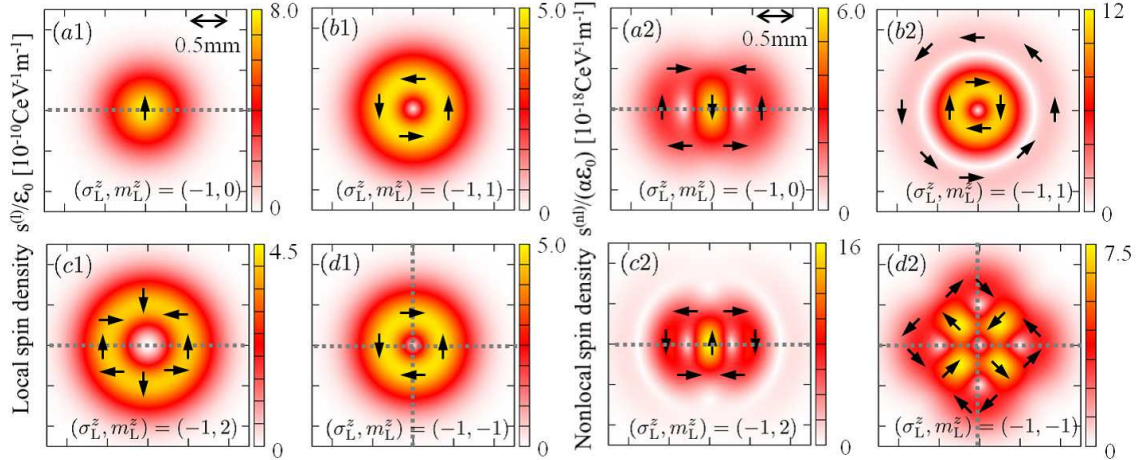}
\caption{(color online) Snapshots of the spin density induced by the electric field of the optical twisted light beam with (a1) and (a2) $ (\sigma^z_{\textrm{L}}, m^z_{\textrm{L}})=(-1,0)$, (b1) and (b2) $ (\sigma^z_{\textrm{L}}, m^z_{\textrm{L}})=(-1,1)$, (c1) and (c2) $ (\sigma^z_{\textrm{L}}, m^z_{\textrm{L}})=(-1,2)$, and (d1) and (d2) $ (\sigma^z_{\textrm{L}}, m^z_{\textrm{L}})=(-1,-1)$.
The left (right) panels show the local (nonlocal) spin density.
The color map and the direction of the arrow show the magnitude and direction of the spin density, respectively. 
The parameters are the same as those in Fig. \ref{fig:fig3}. 
}\label{fig:fig4} 
\end{figure*}

We turn to discuss the spin density induced by the electric field of the twisted light beam in the same setup as that considered in Sec. \ref{sec:5-2}. 
As discussed in Sec. \ref{sec:4-2}, the induced spin density can be divided into the local and nonlocal ones as 
$\bm{s}=\bm{s}^{(\textrm{l})}+\bm{s}^{(\textrm{nl})}$.
The local spin density $\bm{s}^{(\textrm{l})}$ is described in Eq. (\ref{eq:local-spin}) and its snapshots for several $m_{\textrm{L}}^z$ are shown in Figs. \ref{fig:fig4}(a1)--\ref{fig:fig4}(d1).  
The direction of $\bm{s}^{(\textrm{l})}$ is perfectly perpendicular to the electric field.
We find that the dynamical vortex-like spin structure is generated by the twisted light beam and 
the winding number of the local spin density, which is defined by Eq. (\ref{eq:winding-number}) with $\bm{n}=\bm{s}^{(\textrm{l})}$, is identical to that of the electric field, 
\begin{align}
\omega_v[\bm{s}^{(\textrm{l})}]=\omega_v[\bm{E}]=-\sigma_{\textrm{L}}^z m_{\textrm{L}}^z
\end{align}
[see Figs. \ref{fig:fig4}(a1)--\ref{fig:fig4}(d1) and Figs. \ref{fig:fig2}(a)--\ref{fig:fig2}(d)].
On the other hand, the nonlocal spin density $\bm{s}^{(\textrm{nl})}$ is proportional to the spatial gradient of the charge density [see the second term of Eq. (\ref{eq:3-31})], 
and can be estimated by using Eq. (\ref{eq:5-2-3}) as 
$\bm{s}^{(\textrm{nl})} \simeq  \frac{\alpha\ell}{2e} \left( \bm{z} \times \bm{\nabla}\right)  \bar{\rho}_e$.
The snapshots of $\bm{s}^{(\textrm{nl})}$ are shown in Figs. \ref{fig:fig4}(a2)--\ref{fig:fig4}(d2).
We find that dynamical vortex-like spin structures appear and the spin density becomes zero at the center of the vortex. Here, we note that $\bm{s}^{(\textrm{nl})}$ and $\bm{\nabla} \rho_e$ share the same winding number as they are perfectly perpendicular to each other. Since the winding number $w_v(\bm{\nabla}\rho_e$) is $1 (-1)$ around the maxima and minima (the saddle points) of $\rho_e$,  the centers of the spin vortices locate at the extrema (minima, maxima, and saddle points) of $\rho_e$, and therefore, they align on the symmetry axis of the distribution of $\rho_e$ [the dotted lines in Fig. \ref{fig:fig3}]. 
For the cases shown in Figs. \ref{fig:fig4}(a2)--\ref{fig:fig4}(d2), all spin vortices have the winding number $w_v(\bm{s}^{(\textrm{nl})})=1$ except for the one at the center of Fig. \ref{fig:fig4}(d2), which corresponds to the saddle point of $\rho_e$ and has the winding number $-1$. Since $\bm{s}^{(\textrm{nl})}$ is related to $\bm{\nabla} \cdot \bm{E}$ rather than $\bm{E}$, the configuration of the spin vortices depends on the total angular momentum $j_L^z$.
Note however that with the parameters for a realistic system, $|\bm{s}^{\textrm{(nl)}}|/|\bm{s}^{\textrm{(l)}}|$ is in the order of  $\ell^2/d_0^2 \sim 10^{-7}$ and $\bm{s}^{\textrm{(nl)}}$ is negligibly small as compared with $\bm{s}^{\textrm{(l)}}$. 
When we turn off the beam, $\bm{s}^{\textrm{(nl)}}$ becomes prominent and diffusively propagates.
We expect that the photo-induced spin texture can be observed by pump prove technique with the twisted light beam and circularly polarized light beam\cite{rf:Kirilyuk}.

\subsection{Charge and spin currents}\label{sec:5-4} 
\begin{figure*}[htbp]\centering 
\includegraphics[scale=0.3]{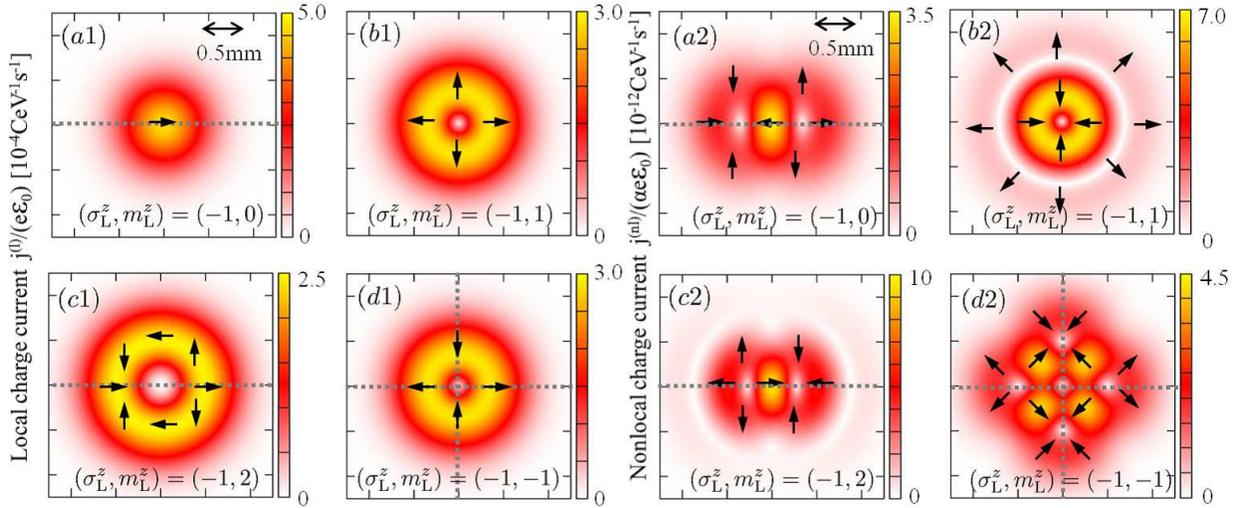}
\caption{(color online) Snapshots of the charge current density induced by the electric field of the optical twisted light beam with (a1) and (a2) $ (\sigma^z_{\textrm{L}}, m^z_{\textrm{L}})=(-1,0)$, (b1) and (b2) $ (\sigma^z_{\textrm{L}}, m^z_{\textrm{L}})=(-1,1)$, (c1) and (c2) $ (\sigma^z_{\textrm{L}}, m^z_{\textrm{L}})=(-1,2)$, and (d1) and (d2) $ (\sigma^z_{\textrm{L}}, m^z_{\textrm{L}})=(-1,-1)$.
The left (right) panels show the local (nonlocal) charge current density.
The color map and the direction of the arrow show the magnitude and direction of the charge current density, respectively. 
The parameters are the same as those in Fig. \ref{fig:fig3}.
}\label{fig:fig5} 
\end{figure*}

The profile of the charge current is similar to that of the spin 
because of the spin momentum locking on the surface of the TI.
In Fig. \ref{fig:fig5}, we show the snapshots of the charge current for various angular momentum of light, where left (right) four panels depict the local (nonlocal) components. Reflecting the relation $\bm{j} \parallel \bm{z}\times \bm s$, the magnitude of $\bm{j}^{(l,nl)}$ has the same profile as that of $|\bm{s}^{(l,nl)}|$, while the direction of $\bm{j}^{(l,nl)}$ is obtained by rotating $\bm{s}^{(l,nl)}$ by $-\pi/2$ about the $z$ axis. As in the case of the spin density, the local (nonlocal) part of the charge current is related to $\bm {E}$ ($\bm{\nabla}\cdot \bm{E}$) and hence its configuration is mainly determined by $m_L^z$ ($j_L^z$).

\begin{figure*}[htbp]\centering 
\includegraphics[scale=0.3]{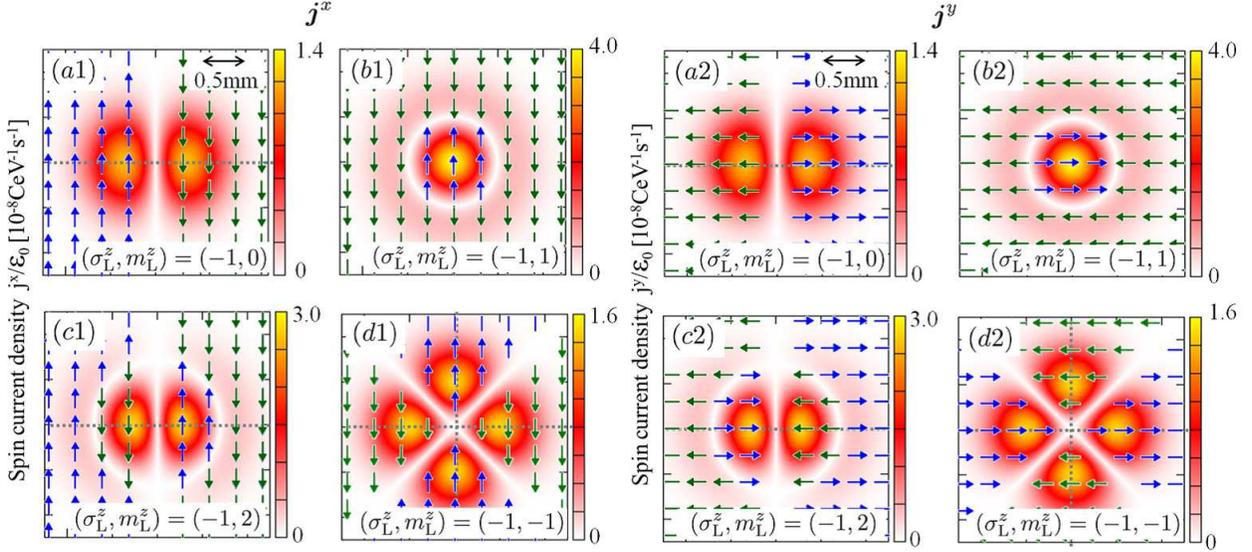}
\caption{(color online) Snapshots of the spin current density induced by the optical twisted light beam with  (a1) and (a2) $ (\sigma^z_{\textrm{L}}, m^z_{\textrm{L}})=(-1,0)$, (b1) and (b2) $ (\sigma^z_{\textrm{L}}, m^z_{\textrm{L}})=(-1,1)$, (c1) and (c2) $ (\sigma^z_{\textrm{L}}, m^z_{\textrm{L}})=(-1,2)$, and (d1) and (d2) $ (\sigma^z_{\textrm{L}}, m^z_{\textrm{L}})=(-1,-1)$.
The left (right) panels show current of the x (y) component of spin.
The color map shows the magnitude of the spin current density.
The blue (green) arrows show the direction of the flow of the $x$ ($y$) component of the spin.
As in the case of the charge density, the distribution of the spin current depends on the total angular momentum of the twisted light beam. 
The parameters are the same as those in Fig. \ref{fig:fig3}.
}\label{fig:fig6}
\end{figure*}

Figure \ref{fig:fig6} shows the light-induced spin current.  
As one can see from Eq. (55), the magnitude of the spin current is proportional to $|\rho_e|$ and the direction of the spin and its current perfectly perpendicular to each other. 
These properties also come from the spin-momentum locking on the surface of the TI.
We find that the spatial profile of the spin current is different from that of the (spin-polarized) charge current, which are shown in Figs. \ref{fig:fig6} and \ref{fig:fig5} respectively. 
In fact, they are related to each other via Eq. (\ref{eq:4-4-4}) and only the nonlocal charge current couples to the spin current.

\section{Conclusion}
We study the charge density, the spin density, the charge current density, and the spin current density induced by a twisted light beam shined on a disordered surface of a doped TI by using the Keldysh-Green's function method.
We have discussed the responses of charge and spin to the space-time dependent electric field. The obtained results are summarized in Tab. I.  The effect of the electric field on the electric charge is twofold. First, it induces a charge current along the direction of the electric field. Second, the inhomogeneity of the electric field causes a gradient of the charge density, which then leads to a diffusive charge current. We call the former the local charge current and the latter the nonlocal charge current, based on whether the charge current depends only on the local electric field or is affected by the nonlocal one. Since the spin and momentum of electrons on a surface of a TI are locked to be perpendicular to each other, the emergence of the charge current implies that the spin density is induced in the perpendicular direction to the charge current. Our calculation based on the linear response theory gives the analytic description for the local and nonlocal spin densities as well as the local and nonlocal charge current densities. 
We also find that the induced charge density also gives rise to a spin current, which is related to the nonlocal part of the charge current via Eq. (\ref{eq:4-4-4}).

By taking into account the spatial and temporal configuration of the electric field associated with a twisted light beam, we have shown that various types of spin vortices arise. Since the local spin density is perpendicular to the electric field, their winding numbers are identical and determined by the product of the spin and orbital angular momentum of the twisted light beam.  In this paper, we have assumed that the time and length scales of the diffusion of electrons is much faster than those of incident light. In such a situation, we can approximate the nonlocal electric field with the bare electric field, and the nonlocal quantities are described using the divergence of the electric field. 
Thus, the configurations of the nonlocal densities and currents are determined by the total angular momentum of the twisted light beam.

\begin{table*}
\caption{Dependence on the applied electric field $\bm{E}$ of the induced charge density $\rho_e$, spin current density $j^\alpha_i$, spin density $\bm{s}$, and charge current density $\bm{j}$ on a disordered surface of a  doped TI, where $\langle \bm{E}\rangle_{\rm D}$ is defined in Eq. (\ref{eq:3-25}). 
The most right column is the winding number of the spin density and the charge current density (which are identical due to the spin-momentum locking). In the distribution of the nonlocal part of $\bm{s}$ and $\bm{j}$, several vortices appear, each of which has the winding number $+1$ or $-1$. The configuration of the vortices is determined by the total angular momentum of light, $j_{\mathrm{L}}^z=\sigma_{\mathrm{L}}^z +m_{\mathrm{L}}^z$. }
\begin{center}
\begin{tabular}{lccccc}
\hline \hline
\ & $\rho_e$  & $j_i^\alpha $ & $\bm{s}$  & $\bm{j}$ &$\omega_v[\bm{s}],\omega_v[\bm{j}]$ \\
\hline
Local \ & -- & --   
& $\bm{z}\times \bm{E}$ 
& $\bm{E}$ &$-\sigma_{\textrm{L}}^z m_{\textrm{L}}^z$   \\
Nonlocal \
& $\bm{\nabla}\cdot \langle \bm{E} \rangle_\textrm{D}$ 
&\ $\epsilon_{z\alpha i} \bm{\nabla}\cdot \langle \bm{E} \rangle_\textrm{D}$ 
&\ $( \bm{z} \times \bm{\nabla} ) \bm{\nabla}\cdot \langle \bm{E} \rangle_\textrm{D}$ 
&\ $\bm{\nabla} (\bm{\nabla}\cdot \langle \bm{E} \rangle_\textrm{D}) $
&\ $\pm1$ 
\\
\hline
\hline
\end{tabular}
\end{center}
\end{table*}

\section{acknowledgment}
The authors would like to thank A. Dutt for valuable discussions.
This work was supported by Grant-in-Aid for Young Scientists (B) (No. 15K17726), by Grants-in-Aid for Scientific Research on Innovative Areas ``Topological Materials Science'' 
(No. 15H05853 and No. 15H05855) from the Ministry of Education, 
Culture, Sports, Science, and Technology, Japan (MEXT), by Grant-in-Aid for Challenging Exploratory Research (No. 15K13498), and by
the Core Research for Evolutional Science and Technology (CREST) of the Japan Science.
K.T. acknowledges support from a Grant-in-Aid for Japan Society for the Promotion of Science (JSPS) Fellows.

\appendix

\section{Derivation of $\hat{I}_{\mu}$}\label{sec:A} 
We estimate $\hat{I}_{\mu}$ in Eq. (\ref{eq:3-14}) by expanding with respect to $\bm{q}$ and $\Omega$ and by using the Green's function  
\begin{align} \notag 
\hat{g}^\textrm{r}_{\bm{k}\pm\frac{\bm{q}}{2}, \pm\frac{\Omega}{2}} 
	= & \hat{g}^\textrm{r}_{\bm{k}} \pm \frac{1}{2}\left[ \sum_{\xi=x,y}q_\xi \frac{\partial \hat{g}^\textrm{r}_{\bm{k}} }{\partial k_\xi} 
		+ \Omega \frac{\partial \hat{g}^\textrm{r}_{\bm{k},\omega} }{\partial \omega}\biggr|_{\omega\to0}  \right]   
	    + \frac{1}{8} \sum_{\xi,\xi'=x,y} q_\xi q_{\xi'} \frac{\partial^2 \hat{g}^\textrm{r}_{\bm{k}} }{\partial k_\xi \partial k_{\xi'}} 
		+O(q^3, q\Omega, \Omega^2), 
\end{align}
where we use the short hand notation $\hat{g}^\textrm{r}_{\bm{k}} \equiv \hat{g}^\textrm{r}_{\bm{k},\omega=0}$. 
$\hat{I}_{\mu}$ is decomposed into four terms as 
\begin{align} \label{eq:coefficient pi-0} 
\hat{I}_\mu   & \equiv  \hat{I}_\mu^{(0)} + \Omega \hat{I}_{\mu}^{(1)}  + \sum_{\xi = x,y}q_\xi \hat{I}_{\mu \xi}^{(2)}  + \sum_{\xi,\xi' = x,y} q_\xi q_{\xi'}  \hat{I}_{\mu\xi \xi' }^{(3)} , 
	\\ \label{eq:coefficient pi-1} 
\hat{I}_\mu^{(0)} 
	& =\sum_{\bm{k}}  \hat{g}^\textrm{r}_{\bm{k}} \hat{\sigma}_\zeta \hat{g}^\textrm{a}_{\bm{k}}, 
	\\ \label{eq:coefficient pi-2} 
 \hat{I}_{\mu}^{(1)} 
	& = \frac{1}{2}\sum_{\bm{k}} \biggl(  \hat{g}^\textrm{r}_{\bm{k}} \hat{\sigma}_\zeta \left.\frac{\partial \hat{g}^\textrm{a}_{\bm{k},\omega} }{\partial \omega} \right|_{\omega\to0}
		- \textrm{h.c}  \biggr),
	\\ \label{eq:coefficient pi-3} 
\hat{I}_{\mu \xi }^{(2)} 
	& = \frac{1}{2} \sum_{\bm{k}} \biggl( \hat{g}^\textrm{r}_{\bm{k}} \hat{\sigma}_\zeta \frac{\partial \hat{g}^\textrm{a}_{\bm{k}} }{\partial k_\xi} - \textrm{h.c} \biggr),
	\\ \label{eq:coefficient pi-4} 
\hat{I}_{\mu \xi \xi'}^{(3)}
	& = \frac{1}{8} \sum_{\bm{k}}  \left( \hat{g}^\textrm{r}_{\bm{k}} \hat{\sigma}_\zeta \frac{\partial^2 \hat{g}^\textrm{a}_{\bm{k}}}{\partial k_\xi \partial k_{\xi'} } 
						+ \frac{\partial^2 \hat{g}^\textrm{r}_{\bm{k}}}{\partial k_\xi \partial k_{\xi'} }  \hat{\sigma}_\zeta \hat{g}^\textrm{a}_{\bm{k}}
						- 2 \frac{\partial \hat{g}^\textrm{r}_{\bm{k}}}{\partial k_\xi}  \hat{\sigma}_\zeta \frac{\partial \hat{g}^\textrm{a}_{\bm{k}}}{\partial k_{\xi'}} 
						 \right).
\end{align}
In Appendix \ref{sec:E}, we list useful formulas for the integral of the Green's functions, which are used in the following calculations.

\subsection{Calculation of $\hat{I}_{\mu=x,y,z}^{(0)} $}
First, to calculate $\hat{I}^{(0)}_{\mu=x,y,z} $ in Eq. (\ref{eq:coefficient pi-1}), we divide $\hat{g}^\textrm{r}_{\bm{k}} \hat{\sigma}_\mu \hat{g}^\textrm{a}_{\bm{k}}$ into the even and odd functions of $\bm{k}$:
\begin{align}
\hat{g}^\textrm{r}_{\bm{k}} \hat{\sigma}_\mu \hat{g}^\textrm{a}_{\bm{k}} = D^\textrm{r} D^\textrm{a} \hat{\mathcal{Q}} \hat{\sigma}_\mu \hat{\mathcal{Q}}^\dag,
\end{align}
where $D^\textrm{r}$, $D^\textrm{a}=[D^\textrm{r}]^*$, $\hat{\mathcal{Q}}$, $\hat{\mathcal{Q}}^\dag$, $h$, and $h^*$ are defined by 
\begin{align}
D^\textrm{r} & \equiv (h^2 - \hbar^2 \tilde{v}_{\rm{F}}^2 k^2 )^{-1}, 
	\\
\hat{\mathcal{Q}}& \equiv  h+ \hbar \tilde{v}_{\textrm{F}} \bm{k}\cdot ( \bm{z} \times \hat{\bm{\sigma}}), 
	\\
\hat{\mathcal{Q}}^\dag& \equiv  h^* + \hbar \tilde{v}_{\textrm{F}} \bm{k}\cdot ( \bm{z} \times \hat{\bm{\sigma}}), 
 	\\
	h & \equiv \epsilon_{\textrm{F}} +i\eta, 
	\\
	h^* & \equiv \epsilon_{\textrm{F}} - i\eta.
\end{align}
$D^\textrm{r}$ and $D^\textrm{a}$ are the even functions of $\bm{k}$. 
$\hat{\mathcal{Q}} \hat{\sigma}_\mu \hat{\mathcal{Q}}^\dag$ is represented by 
\begin{align}\notag
\hat{\mathcal{Q}} \hat{\sigma}_\mu \hat{\mathcal{Q}}^\dag 
	& = \left[ |h|^2 \hat{\sigma}_\mu +\hbar^2 \tilde{v}_{\textrm{F}}^2\sum_{\ell,\ell'=x,y} k_{\ell} k_{\ell'} ( \bm{z} \times \hat{\bm{\sigma}})_\ell \hat{\sigma}_\mu ( \bm{z} \times \hat{\bm{\sigma}})_{\ell'} \right] 
		\\ \label{eq:app-12} 
		&\hspace{10mm}
		+ \hbar \tilde{v}_{\textrm{F}} \sum_{\ell=x,y}\left[ h k_\ell \hat{\sigma}_\mu ( \bm{z} \times \hat{\bm{\sigma}})_\ell  + h^* k_\ell  ( \bm{z} \times \hat{\bm{\sigma}})_\ell \hat{\sigma}_\mu \right].
\end{align}
The first and second terms are corresponding to the even and odd functions of $\bm{k}$, respectively. 
In the following calculation, we simply assume that the surface of the TI is isotropic as a function of $\bm{k}$: 
$k_x^2 = k_y^2 = k^2/2$. 
By taking an average over the direction of $\bm{k}$, $\langle \ \rangle_k$, in Eq. (\ref{eq:app-12}) and using  $\langle k_\ell \rangle_k=0$ and $\langle k_\ell k_{\ell'}\rangle_k  = \frac{1}{2} k^2 \delta_{\ell \ell'}$, we obtain
\begin{align}  \label{eq:app-13} 
\langle \hat{\mathcal{Q}} \hat{\sigma}_\mu \hat{\mathcal{Q}}^\dag \rangle_k 
	& = |h|^2 \hat{\sigma}_\mu 
	   + \frac{1}{2} \hbar^2 \tilde{v}_{\textrm{F}}^2 k^2 \sum_{\ell=x,y} ( \bm{z} \times \hat{\bm{\sigma}})_\ell \hat{\sigma}_\mu ( \bm{z} \times \hat{\bm{\sigma}})_{\ell}.
\end{align}
The second term in the above equation becomes  
\begin{align}  \label{eq:app-14} 
 \sum_{\ell =x,y } (\bm{z} \times \hat{\bm{\sigma}} )_\ell \hat{\sigma}_\mu (\bm{z} \times \hat{\bm{\sigma}} )_{\ell}
	 = & \sum_{\ell = x,y } (2\delta_{\ell \mu} \hat{\sigma}_\ell - \delta_{\ell\ell} \hat{\sigma}_\mu)
	\\
	 = & -2\delta_{\mu z}\hat{\sigma}_z.
\end{align}
Thus, we obtain $\hat{I}_\mu^{(0)}$ as   
\begin{align} 	\label{eq:app-15} 
\hat{I}_{\mu=x,y}^{(0)} 
	& = |h|^2\sum_{\bm{k}}   |D^\textrm{r}|^2 \hat{\sigma}_\mu, 
	\\ 
	\label{eq:app-16} 
\hat{I}_{z}^{(0)}
	& = \sum_{\bm{k}} (|h|^2 - \hbar^2 \tilde{v}_{\textrm{F}}^2 k^2)  |D^\textrm{r}|^2 \hat{\sigma}_z. 			
\end{align}
The integral of Eq. (\ref{eq:app-15}) is obtained by  
\begin{align} \notag
 \frac{1}{V} \sum_{\bm{k}}   |D^\textrm{r}|^2  
 =& \frac{1}{2\pi} \int_0^\infty   \frac{kdk}{ [h^2 - \hbar^2 \tilde{v}_{\textrm{F}}^2 k^2 ] [ (h^*)^2 - \hbar^2 \tilde{v}_{\textrm{F}}^2 k^2 ]} 
	= \frac{\nu_e }{2\epsilon_{\textrm{F}}} \int_0^\infty \frac{dx}{ [h^2 -x ] [(h^*)^2 -x ]} 
		\\ 
	=& \frac{\nu_e }{2\epsilon_{\textrm{F}}[ h^2 - (h^*)^2 ]} \int^\infty_0 dx \biggl[ \frac{1}{ x -h^2 } - \frac{1}{ x - (h^*)^2 }  \biggr],
\end{align}
where 
$\nu_e =  \frac{ \epsilon_{\textrm{F}}}{2\pi \hbar^2 \tilde{v}_{\textrm{F}}^2 }$ is the density of states on the surface. 
Here, the above integral is given by 
\begin{align} \label{eq:app-18} 
  \int_0^\infty dx \biggl[ \frac{1}{ x -h^2 } - \frac{1}{ x - (h^*)^2 }  \biggr ]
	 =  \log{ \biggl| \frac{x-h^2}{x-(h^*)^2} \biggr|_{x\to0}^{x\to\infty} }
	 =   i \biggl[ \arg{(x -h^2)} - \arg{(x -(h^*)^2)} \biggr]_{x\to0}^{x\to\infty}.
\end{align}
In the above equation, we have used $\log{z} = {\textrm{Log}}{|z|} + i \arg z$, where $z=a+ib = |z| e^{i\theta}$, $a, b \in \textrm{R}$.
Here, $\theta=\arg {z}$ is 
\begin{align}
\theta
	=&\left\{
\begin{array}{ll}
			\tan^{-1}(b/a)   &(a>0 \textrm{ and } b>0)\\ 
		\theta = \pi + \tan^{-1}(b/a) &(a<0)\\
		\theta = 2\pi + \tan^{-1}(b/a) & (a>0 \textrm{ and } b<0)
\end{array}\right..
\end{align}
Thus,  $\arg{(x -h^2)}|_{x\to0}^{x\to\infty} = \pi +o(\hbar/\epsilon_{\textrm{F}}\tau)$, 
$\arg{[x -(h^*)^2]} |_{x\to0}^{x\to\infty} =- \pi +o(\hbar/\epsilon_{\textrm{F}}\tau)) $, and 
\begin{align}
\log \biggl| \frac{x-h^2}{x-(h^*)^2} \biggr|_{x\to0}^{x\to\infty} \simeq 2i\pi
\end{align}
are satisfied. 
We have 
$\frac{1}{L^2} \sum_{\bm{k}}   |D^\textrm{r}|^2  =  \frac{ \pi  \nu_e }{ 4 \eta \epsilon^2_{\textrm{F}} } + o(\hbar/\epsilon_{\textrm{F}}\tau)$ 
and 
\begin{align} 	\label{eq:coefficient c-1 no3} 
\hat{I}^{(1)}_{\mu=x,y}  
	& =\frac{ \pi  \nu_e }{ 4 \eta } \hat{\sigma}_\mu. 
\end{align}
Eq. (\ref{eq:app-16}) can be estimated around the Fermi energy $k \to k_{\textrm{F}} \equiv  \epsilon_{\textrm{F}}/(\hbar \tilde{v}_{\textrm{F}})$ as
\begin{align}
\hat{I}_{z}^{(0)}
	& \sim  (|h|^2 - \hbar^2 \tilde{v}_{\textrm{F}}^2 k_{\textrm{F}}^2) \sum_{\bm{k}} |D^\textrm{r}|^2 \hat{\sigma}_z
		=o\left(\frac{\hbar}{\epsilon_{\textrm{F}}\tau}\right).
\end{align}

\subsection{Calculation of $\hat{I}^{(1)}_{\mu=x,y,z}$}
To calculate $\hat{I}^{(1)}_{\mu=x,y }$  in Eq. (\ref{eq:coefficient pi-2}), we divide  
$ \hat{g}^\textrm{r}_{\bm{k}} \hat{\sigma}_\mu \frac{\partial \hat{g}^\textrm{a}_{\bm{k},\omega}}{\partial \omega}|_{\omega\to0} = -\hbar \hat{g}^\textrm{r}_{\bm{k}}  \hat{\sigma}_\mu (\hat{g}^\textrm{a}_{\bm{k}} )^2 
	= -\hbar D^\textrm{r} (D^\textrm{a})^2 \hat{\mathcal{Q}} \hat{\sigma}_\mu (\hat{\mathcal{Q}}^\dag)^2$ 
	into the even and odd functions of $\bm{k}$.
Here, $(\hat{\mathcal{Q}}^\dag)^2 $ and $\hat{\mathcal{Q}} \hat{\sigma}_\mu(\hat{\mathcal{Q}}^\dag)^2$ becomes 
\begin{align}  
(\hat{\mathcal{Q}}^\dag)^2 
	= &  [ (h^*)^2 + \hbar^2 \tilde{v}_{\textrm{F}}^2 k^2 ] + 2 h^*  \hbar \tilde{v}_{\textrm{F}} k_\ell  (\bm{z} \times \hat{\bm{\sigma}})_\ell,  
\\ \notag
\hat{\mathcal{Q}} \hat{\sigma}_\mu(\hat{\mathcal{Q}}^\dag)^2
	 = & \biggl[ h \{(h^*)^2 + \hbar^2 \tilde{v}_{\textrm{F}}^2 k^2  \} \hat{\sigma}_\mu  
	  + 2 h^* \hbar^2 \tilde{v}_{\textrm{F}}^2 k_\ell k_{\ell'} (\bm{z}\times \hat{\bm{\sigma}})_\ell  \hat{\sigma}_\mu (\bm{z}\times \hat{\bm{\sigma}})_{\ell'}   \biggr]
	\\ 
		& + \biggl[ 2 |h|^2 \hbar \tilde{v}_{\textrm{F}}   k_\ell (\bm{z}\times \hat{\bm{\sigma}})_\ell \hat{\sigma}_\mu \hat{\sigma}_\mu  
		 + \{ (h^*)^2 + \hbar^2 \tilde{v}_{\textrm{F}}^2 k^2 \} \hbar \tilde{v}_{\textrm{F}} k_\ell (\bm{z}\times \hat{\bm{\sigma}})_\ell   \hat{\sigma}_\mu 
			\biggr].
\end{align}
The first and second terms of the above equations are the even and odd functions of $\bm{k}$ .
Then, we have
\begin{align} 
\langle \hat{\mathcal{Q}} \hat{\sigma}_{\mu}(\hat{\mathcal{Q}}^\dag)^2 \rangle_k 
	 = &  h \{(h^*)^2 + \hbar^2 \tilde{v}_{\textrm{F}}^2 k^2  \} \hat{\sigma}_\mu \ \ \ (\mu=x,y), 
	 \\
\langle \hat{\mathcal{Q}} \hat{\sigma}_{z}(\hat{\mathcal{Q}}^\dag)^2 \rangle_k 
	 = &  h \{(h^*)^2 + \hbar^2 \tilde{v}_{\textrm{F}}^2 k^2  \} \hat{\sigma}_z
	    -  2 h^* \hbar^2 \tilde{v}_{\textrm{F}}^2 k^2 \hat{\sigma}_z \ \ \ (\mu=z),
\end{align}
from which
$\hat{I}^{(1)}_{\mu} = -\frac{\hbar}{2} \sum_{\bm{k}} [ \hat{g}^\textrm{r}_{\bm{k}} \hat{\sigma}_\mu (\hat{g}^\textrm{a}_{\bm{k}} )^2- \textrm{h.c.}  ]$ is obtained by
\begin{align} \notag 
\hat{I}^{(1)}_{\mu=x,y}  
	& =-\frac{\hbar}{2}   \biggl\{ h \sum_{k}   D^\textrm{r} (D^\textrm{a})^2 
		\bigl[  (h^*)^2 + \hbar^2 \tilde{v}_{\textrm{F}}^2 k^2   \bigr] \hat{\sigma}_\mu -\textrm{h.c.} \biggr\}
	\\ \notag 
	& =-\frac{\hbar}{2}   \biggl[   \frac{\nu_e}{8\eta^2} 
		 \left( i\pi - \frac{2\eta^2}{\epsilon_{\textrm{F}}^2 }  \right)  \left(1+i\frac{\eta}{\epsilon_{\textrm{F}}} \right) 
		  -\textrm{c.c.} \biggr] \hat{\sigma}_\mu + o\left(\frac{\hbar}{\epsilon_{\textrm{F}}\tau}\right)
	\\ \label{eq:coefficient c-2 no2} 
	&  =   \frac{-i\hbar  \pi \nu_e}{8\eta^2 }  \hat{\sigma}_\mu + o\left(\frac{\hbar}{\epsilon_{\textrm{F}}\tau}\right),
	\\\notag
\hat{I}^{(1)}_{z}  
	& =-\frac{\hbar}{2}   \biggl\{ \sum_{k}   D^\textrm{r} (D^\textrm{a})^2 
		\bigl[  h(h^*)^2 + (h - 2h^*)\hbar^2 \tilde{v}_{\textrm{F}}^2 k^2   \bigr] \hat{\sigma}_\mu -\textrm{h.c.} \biggr\}
	\\ \notag 
	& =-\frac{\hbar}{2}\left[\frac{-i\pi\nu_e}{4\epsilon_\textrm{F}^2}\left(1+i\frac{\eta}{\epsilon_{\textrm{F}}} \right) - \textrm{c.c.}\right]\hat{\sigma}_z + o\left(\frac{\hbar}{\epsilon_{\textrm{F}}\tau}\right)
	\\ \label{eq:coefficient c-2 no3} 
	&  =   \frac{i\hbar  \pi \nu_e}{4\epsilon_\textrm{F}^2 }  \hat{\sigma}_\mu + o\left(\frac{\hbar}{\epsilon_{\textrm{F}}\tau}\right).
\end{align}
%

\subsection{Calculation of $\hat{I}^{(2)}_{\mu(=x,y,z)\xi } $ }
$\frac{\partial \hat{g}^\textrm{r}_{\bm{k}}}{\partial k_\xi} $ is given by 
\begin{align}
\frac{\partial \hat{g}^\textrm{r}_{\bm{k}}}{\partial k_\xi} 
		& =  \hbar \tilde{v}_{\textrm{F}} (\bm{z}\times \hat{\bm{\sigma}})_\xi D^\textrm{r} + 2\hbar^2\tilde{v}_{\textrm{F}}^2 k_\xi \hat{\mathcal{Q}} (D^\textrm{r})^2.
	 \label{eq:differential1 gr-0} 
\end{align}
By using Eq. (\ref{eq:differential1 gr-0}), 
$\langle \frac{\partial \hat{g}^\textrm{r}_{\bm{k}}}{\partial k_\xi} \hat{\sigma}_\mu \hat{g}^\textrm{a}_{\bm{k}} \rangle_k$ becomes
\begin{align} \notag 
\langle \frac{\partial \hat{g}^\textrm{r}_{\bm{k}}}{\partial k_\xi} \hat{\sigma}_\mu \hat{g}^\textrm{a}_{\bm{k}} \rangle_k
	= &  \left[  \hbar \tilde{v}_{\textrm{F}} (\bm{z}\times \hat{\bm{\sigma}})_\xi D^\textrm{r} + 2\hbar^2\tilde{v}_{\textrm{F}}^2 k_\xi \hat{\mathcal{Q}} (D^\textrm{r})^2  
	     \right]  
	      \hat{\sigma}_\mu D^\textrm{a} \hat{\mathcal{Q}}^\dag 
	\\
	 = &   \hbar \tilde{v}_{\textrm{F}} (\bm{z}\times \hat{\bm{\sigma}})_\xi \hat{\sigma}_\mu  \langle  \hat{\mathcal{Q}}^\dag\rangle_k |D^\textrm{r}|^2 			
	    + 2\hbar^2\tilde{v}_{\textrm{F}}^2 \langle  k_\xi \hat{\mathcal{Q}} \hat{\sigma}_\mu  \hat{\mathcal{Q}}^\dag \rangle_k (D^\textrm{r})^2 D^\textrm{a}. 
\end{align}
Here, $\langle  \hat{\mathcal{Q}}^\dag\rangle_k$ and $\langle  k_\xi \hat{\mathcal{Q}} \hat{\sigma}_\mu  \hat{\mathcal{Q}}^\dag \rangle_k$ are given by
\begin{align}
 \langle  \hat{\mathcal{Q}}^\dag \rangle_k 
	& = h^* , 
	\\ \notag 
 \langle  k_\xi \hat{\mathcal{Q}} \hat{\sigma}_\mu \hat{\mathcal{Q}}^\dag \rangle_k
	& = \hbar \tilde{v}_{\textrm{F}} \langle  k_\xi  k_\ell \rangle_k \{ h \hat{\sigma}_\mu (\bm{z} \times \hat{\bm{\sigma}})_\ell + h^*  (\bm{z} \times \hat{\bm{\sigma}})_\ell \hat{\sigma}_\mu \} 
	\\
	& =  \frac{\hbar \tilde{v}_{\textrm{F}} }{2} k^2 \epsilon_{\xi z u} ( h \hat{\sigma}_\mu \hat{\sigma}_u + h^*\hat{\sigma}_u \hat{\sigma}_\mu ).
\end{align}
$\hat{I}^{(2)}_{\mu \xi}$ in Eq. (\ref{eq:coefficient pi-3}) is given by  
\begin{align} \label{eq:differential-1 grgrga 2} 
\hat{I}^{(2)}_{\mu \xi} 
	= & \frac{\hbar \tilde{v}_{\textrm{F}}}{2} \epsilon_{\xi z u} \sum_k 
		\biggl[    |D^\textrm{r}|^2 (  h \hat{\sigma}_\mu \hat{\sigma}_u -h^* \hat{\sigma}_u \hat{\sigma}_\mu)   
          +  \hbar^2\tilde{v}_{\textrm{F}}^2 k^2 |D^\textrm{r}| ( D^\textrm{a} - D^\textrm{r})   ( h \hat{\sigma}_\mu \hat{\sigma}_u + h^*\hat{\sigma}_u \hat{\sigma}_\mu )
		\biggr].
\end{align}
Here, $h \hat{\sigma}_\mu \hat{\sigma}_u \pm h^* \hat{\sigma}_u \hat{\sigma}_\mu$ can be transformed as 
\begin{align} 
 h \hat{\sigma}_\mu \hat{\sigma}_u -h^* \hat{\sigma}_u \hat{\sigma}_\mu
	& = 2i\eta \delta_{\mu u} + 2 i \epsilon_{\textrm{F}} \epsilon_{\mu u\nu} \hat{\sigma}_\nu, 
\\
 h \hat{\sigma}_\mu \hat{\sigma}_u + h^* \hat{\sigma}_u \hat{\sigma}_\mu
	& = 2\epsilon_{\textrm{F}}  \delta_{\mu u} - 2  \eta \epsilon_{\mu u\nu} \hat{\sigma}_\nu. 
\end{align}
As a result, $\hat{I}^{(2)}_{\mu \xi}$ is obtained by 
\begin{align}\notag
\hat{I}^{(2)}_{\mu(=x,y) \xi} 
	& =  \frac{\hbar \tilde{v}_{\textrm{F}} i\pi\nu_e }{8 \eta^2} \epsilon_{\xi z u}
		\biggl[   \left( 1+ 2\frac{\eta^2 }{\epsilon_{\textrm{F}}^2} \right) \delta_{\mu u} 
			+ \frac{\eta}{\epsilon_{\textrm{F}}}  \epsilon_{\mu u\nu} \hat{\sigma}_\nu   
		\biggr] +o\left( \frac{\hbar}{\epsilon_\textrm{F} \tau}\right)
	\\	 \label{eq:A29}
	& = \frac{ i\pi \nu_e }{8\eta^2}  \hbar \tilde{v}_{\textrm{F}} \epsilon_{\mu\xi z} +o\left( \frac{\hbar}{\epsilon_\textrm{F} \tau}\right) ,
	\\ \notag
\hat{I}^{(2)}_{z \xi} 
	& =  \frac{\hbar \tilde{v}_{\textrm{F}} i\pi\nu_e }{8 \eta^2} \epsilon_{\xi z u}
		\biggl[   \left( 1+ 2\frac{\eta^2 }{\epsilon_{\textrm{F}}^2} \right) \delta_{z u} 
			+ \frac{\eta}{\epsilon_{\textrm{F}}}  \epsilon_{z u\nu} \hat{\sigma}_\nu   
		\biggr] +o\left( \frac{\hbar}{\epsilon_\textrm{F} \tau}\right)
	\\	\label{eq:A29-z}
	& = \frac{ i\pi \nu_e }{8\epsilon_\textrm{F}\eta}  \hbar \tilde{\xi}_{\textrm{F}} +o\left( \frac{\hbar}{\epsilon_\textrm{F} \tau}\right) .
\end{align}

\subsection{Calculation of $\hat{I}^{(3)}_{\mu(=x,y,z) \xi {\xi'}} $ }
$\hat{I}^{(3)}_{\mu \xi {\xi'}} $ in Eq. (\ref{eq:coefficient pi-4}) is represented by using the partial integral as  
\begin{align}	\label{eq:differential-4 grgrgrga 1} 
\hat{I}^{(3)}_{\mu \xi {\xi'}} 
	 = \frac{1}{4} \sum_{{\bm{k}}}  \left[ \frac{\partial^2 \hat{g}^\textrm{r}_{\bm{k}}}{\partial k_\xi \partial k_{\xi'} }  \hat{\sigma}_\mu \hat{g}^\textrm{a}_{\bm{k}} + \textrm{h.c.} \right].
\end{align}
In order to consider $\langle \frac{\partial^2 \hat{g}^\textrm{r}_{\bm{k}}}{\partial k_\xi \partial k_{\xi'} }  \hat{\sigma}_\mu \hat{g}^\textrm{a}_{\bm{k}} \rangle_k $, we use the following equations:  
\begin{align}
\frac{\partial D^\textrm{r} }{\partial k_{\xi'}} & = 2\hbar^2 \tilde{v}_{\textrm{F}}^2 {k_\xi'} (D^\textrm{r})^2, 
	\\
\frac{\partial \hat{\mathcal{Q}} (D^\textrm{r})^2 }{\partial k_{\xi'}}  
		& = \hbar \tilde{v}_{\textrm{F}} (\bm{z} \times \hat{\bm{\sigma}})_{\xi'}  (D^\textrm{r})^2
			+ 4 \hbar^2 \tilde{v}_{\textrm{F}}^2 k_{\xi'} \hat{\mathcal{Q}}   (D^\textrm{r})^3, 
	\\ 
\frac{\partial^2 \hat{g}^\textrm{r}_{\bm{k}}}{\partial k_\xi \partial k_{\xi'} }
		& =   2\hbar^2 \tilde{v}_{\textrm{F}}^2 
			\biggl\{
			\delta_{\xi\xi'} \hat{\mathcal{Q}} (D^\textrm{r})^2 
				+ 4\hbar^2 \tilde{v}_{\textrm{F}}^2 k_\xi k_{\xi'} \hat{\mathcal{Q}} (D^\textrm{r})^3
				+  \hbar \tilde{v}_{\textrm{F}} [  k_\xi (\bm{z} \times \hat{\bm{\sigma}})_{\xi'}  
									+ k_{\xi'} (\bm{z} \times \hat{\bm{\sigma}})_{\xi} ]   (D^\textrm{r})^2
			\biggr\},
	\\ \notag
\frac{\partial^2 \hat{g}^\textrm{r}_{\bm{k}}}{\partial k_\xi \partial k_{\xi'} } \hat{\sigma}_\mu \hat{g}^\textrm{a}_{\bm{k}}
	 & =   2\hbar^2 \tilde{v}_{\textrm{F}}^2 
			\biggl\{
			\delta_{\xi\xi'} \hat{\mathcal{Q}} (D^\textrm{r})^2\hat{\sigma}_\mu \hat{\mathcal{Q}}^\dag D^\textrm{a} 
					+ 4\hbar^2 \tilde{v}_{\textrm{F}}^2 k_\xi k_{\xi'} \hat{\mathcal{Q}} (D^\textrm{r})^3\hat{\sigma}_\mu \hat{\mathcal{Q}}^\dag D^\textrm{a}
		\\&\hspace{20mm}				
 +  \hbar \tilde{v}_{\textrm{F}} [  k_\xi (\bm{z} \times \hat{\bm{\sigma}})_{\xi'}  
									+ k_{\xi'} (\bm{z} \times \hat{\bm{\sigma}})_{\xi} ]   (D^\textrm{r})^2\hat{\sigma}_\mu \hat{\mathcal{Q}}^\dag D^\textrm{a}
			\biggr\}.
\end{align}
The average of $\bm{k}$ in $\hat{\mathcal{Q}} \hat{\sigma}_\mu \hat{\mathcal{Q}}^\dag$, $k_\xi (\bm{z} \times \hat{\bm{\sigma}})_{\xi'}  \hat{\sigma}_\mu \hat{\mathcal{Q}}^\dag$ and $k_\xi k_{\xi'} \hat{\mathcal{Q}}  \hat{\sigma}_\mu \hat{\mathcal{Q}}^\dag$ become  
\begin{align} \label{eq:differential-4 grgrgrga 20} 
 \langle \hat{\mathcal{Q}} \hat{\sigma}_\mu \hat{\mathcal{Q}}^\dag  \rangle_k 
	 &  = |h|^2 \hat{\sigma}_\mu,
	\\ \label{eq:differential-4 grgrgrga 22} 
 \langle k_\xi (\bm{z} \times \hat{\bm{\sigma}})_{\xi'}  \hat{\sigma}_\mu \hat{\mathcal{Q}}^\dag \rangle_k 
	 & = \frac{1}{2} \hbar \tilde{v}_{\textrm{F}} k^2  (\bm{z} \times \hat{\bm{\sigma}})_{\xi'} \hat{\sigma}_\mu (\bm{z} \times \hat{\bm{\sigma}})_{\xi},
	\\ 
 \langle k_\xi k_{\xi'} \hat{\mathcal{Q}}  \hat{\sigma}_\mu \hat{\mathcal{Q}}^\dag \rangle_k 
	& =  |h|^2 \langle k_\xi k_{\xi'} \rangle_k \hat{\sigma}_\mu 
		 + \hbar^2 \tilde{v}_{\textrm{F}}^2 \langle k_\xi k_{\xi'} k_\ell k_{\ell'} \rangle_k  (\bm{z} \times \hat{\bm{\sigma}})_\ell \hat{\sigma}_\mu (\bm{z} \times \hat{\bm{\sigma}})_{\ell'}.
\end{align}
Here, we have used $\langle k_\xi k_{\xi'} k_\ell k_{\ell'} \rangle_k  = \frac{k^4}{8} (\delta_{\xi \xi'} \delta_{\ell \ell'} +\delta_{\xi \ell} \delta_{\xi' \ell'} +\delta_{\xi \ell'} \delta_{\xi' \ell})$.
Then, $\hat{I}^{(3)}_{\mu \xi {\xi'}}$ is given by  
\begin{align}   \notag 
\hat{I}^{(3)}_{\mu \xi {\xi'}} 
	 =  \frac{\hbar^2 \tilde{v}_{\textrm{F}}^2 }{2} 
		\sum_{\bm{k}} 
		& \biggl\{ \biggl[ 
			\delta_{\xi\xi'} \hat{\sigma}_\mu |h|^2 (D^\textrm{r})^2 D^\textrm{a}
		 + \frac{1}{2} \hbar^2 \tilde{v}_{\textrm{F}}^2 k^2  (D^\textrm{r})^2 D^\textrm{a}
					[  (\bm{z} \times \hat{\bm{\sigma}})_{\xi'} \hat{\sigma}_\mu (\bm{z} \times \hat{\bm{\sigma}})_{\xi}  +   ( \xi \leftrightarrow \xi') ]   
		\\ \notag 
			 &+  
			  	 2 |h|^2 \hbar^2 \tilde{v}_{\textrm{F}}^2 k^2  \delta_{\xi \xi'} \hat{\sigma}_\mu (D^\textrm{r})^3 D^\textrm{a} 
		\\ \notag 
			  &+  \frac{1}{2} \hbar^4 \tilde{v}_{\textrm{F}}^4 k^4   (D^\textrm{r})^3 D^\textrm{a} 
		[  (\bm{z} \times \hat{\bm{\sigma}})_\xi \hat{\sigma}_\mu (\bm{z} \times \hat{\bm{\sigma}})_{\xi'}
			+  ( \xi \leftrightarrow \xi')
		]
			\biggr] 
		\\ 
			& +  \textrm{h.c.} \biggl\}
	\\ \notag 
	 = \frac{\hbar^2 \tilde{v}_{\textrm{F}}^2 }{2} 
		\sum_{\bm{k}} 
		& \biggl[ 
			|h|^2 \delta_{\xi\xi'} \hat{\sigma}_\mu  \biggl(  |D^\textrm{r}|^2 ( D^\textrm{r} + D^\textrm{a} ) 
				+ 2 \hbar^2\tilde{v}_{\textrm{F}}^2 k^2 |D^\textrm{r}|^2 \{ (D^\textrm{r})^2 + (D^\textrm{a})^2 \} \biggr) 
	\\  \notag 
			   & + \biggl[ \hbar^2\tilde{v}_{\textrm{F}}^2 k^2 |D^\textrm{r}|^2 ( D^\textrm{r} + D^\textrm{a} )   
			      + \hbar^4 \tilde{v}_{\textrm{F}}^4 k^4 |D^\textrm{r}|^2 \{ (D^\textrm{r})^2 + (D^\textrm{a})^2 \} \biggr]
\\ \label{eq:differential-4 grgrgrga 4} 
			    & \  \times \bigl[  \epsilon_{\xi z \mu} (\bm{z}\times \hat{\bm{\sigma}})_{\xi'} 
			+ \epsilon_{\xi' z \mu} (\bm{z}\times \hat{\bm{\sigma}})_\xi - \delta_{\xi \xi'} \hat{\sigma}_\mu \bigr]   
			\biggr] .
\end{align}
Here, we have used  
\begin{align} \notag 
 (\bm{z} \times \hat{\bm{\sigma}})_\xi \hat{\sigma}_\mu (\bm{z} \times \hat{\bm{\sigma}})_{\xi'}
			+  ( \xi \leftrightarrow \xi')
	 =&   2 \bigl[ 
		\epsilon_{\xi z \mu} (\bm{z}\times \hat{\bm{\sigma}})_{\xi'} 
			+ \epsilon_{\xi' z \mu} (\bm{z}\times \hat{\bm{\sigma}})_\xi - \delta_{\xi \xi'} \hat{\sigma}_\mu
		\bigr].
\end{align}
 $\hat{I}^{(3)}_{\mu \xi\xi'}$ is given by using equations in Appendix \ref{sec:E} as 
\begin{align}\label{eq:A38}
\hat{I}^{(3)}_{\mu(=x,y) \xi\xi'}
		 = & - \hbar^2 \tilde{v}_{\textrm{F}}^2  \frac{\pi \nu_e }{64\eta^3 } 
			   \bigl[   \epsilon_{\xi z \mu} \epsilon_{\xi' z \nu}  
			+ \epsilon_{\xi' z \mu}\epsilon_{\xi z \nu} + \delta_{\xi \xi'} \delta_{\nu \mu} \bigr]   \hat{\sigma}_\nu +o\left( \frac{\hbar}{\epsilon_\textrm{F} \tau} \right),
	\\\label{eq:A38-z}
\hat{I}^{(3)}_{z \xi\xi'}
		= & \hbar^2 \tilde{v}_{\textrm{F}}^2  \frac{\pi \nu_e }{8\epsilon_\textrm{F}^2\eta } + o\left( \frac{\hbar}{\epsilon_\textrm{F} \tau} \right).
\end{align}

Using $q_\alpha q_\beta \bigl[   \epsilon_{\alpha z \mu} \epsilon_{\beta z \nu}  
			+ \epsilon_{\beta z \mu}\epsilon_{\alpha z \nu} + \delta_{\alpha \beta } \delta_{\nu \mu} \bigr]  
	  = 
	  q_\alpha q_\beta \bigl[   3   \delta_{\alpha \beta } \delta_{\nu \mu} - 2 \delta_{\alpha \nu } \delta_{\beta \mu} \bigr]$,
we obtain 
\begin{align}
q_\alpha q_\beta \hat{I}^{(3)}_{\mu \alpha \beta }
	= - \hbar^2 \tilde{v}_{\textrm{F}}^2 & \frac{\pi \nu_e }{64\eta^3 } 
			  q_\alpha q_\beta \bigl[   3   \delta_{\alpha \beta } \delta_{\nu \mu} - 2 \delta_{\alpha \nu } \delta_{\beta \mu} \bigr]   \hat{\sigma}_\nu.
\end{align}

Thus, $\hat{I}_\mu = \hat{I}_\mu^{(0)} + \Omega \hat{I}_{\mu}^{(1)} 
		+ q_\xi \hat{I}_{\mu \xi}^{(2)} 
	  	  + q_\xi  q_{\xi'} \hat{I}_{\mu \xi \xi' }^{(3)}$
is obtained by 
\begin{align} \notag 
\hat{I}_{\mu=x,y}
		& \simeq   \frac{\pi  \nu_e }{4\eta } 
		  \biggl[ (1 - i \Omega \tau  - \frac{3}{2}D\tau q^2 )  \hat{\sigma}_\mu
		+  i q_\alpha \ell   \epsilon_{\mu \alpha z}
	  	    +D\tau   q_{\mu}  q_\nu \hat{\sigma}_\nu
		\biggr],
	\\
\hat{I}_{\mu=z}
		&  \simeq o\left( \frac{\hbar}{\epsilon_\textrm{F} \tau} \right),
\end{align}
where $D=\frac{1}{2}\tilde{v}_{\textrm{F}}^2\tau$ and $\ell=\tilde{v}_{\textrm{F}}\tau$ are the diffusion constant and the mean free path of the surface electrons, respectively.

\section{Calculation of $\hat{I}_{0}$}\label{sec:B}
We will calculate $\hat{I}_{0}$ using the same formalism in the Appendix \ref{sec:A}. 
Here, $\hat{I}_{0} = \sum_{\bm{k}}  \hat{g}^\textrm{r}_{\bm{k}-\frac{\bm{q}}{2}, -\frac{\Omega}{2}} \hat{g}^\textrm{a}_{\bm{k}+\frac{\bm{q}}{2}, \frac{\Omega}{2}}$ can be expanded with respect to $\bm{q}$ and $\Omega$ within $q\ell \ll 1$ and $\Omega \tau \ll 1$ as
\begin{align} 
\hat{I}_0  & \equiv  \hat{I}_0^{(0)} + \Omega \hat{I}_{0}^{(1)}  + \sum_{\xi=x,y}q_\xi \hat{I}_{0 \xi}^{(2)}  + \sum_{\xi,\xi'=x,y}q_\xi q_{\xi'}  \hat{I}_{0 \xi \xi'  }^{(3)} , 
	\\ \label{eq:coefficient pi-1'} 
\hat{I}_0^{(0)} 
	& =\sum_{\bm{k}}  \hat{g}^\textrm{r}_{\bm{k}}  \hat{g}^\textrm{a}_{\bm{k}}, 
	\\ \label{eq:coefficient pi-2'} 
 \hat{I}_{0}^{(1)} 
	& = \frac{1}{2}\sum_{\bm{k}}   \left(  \hat{g}^\textrm{r}_{\bm{k}}  \left.\frac{\partial \hat{g}^\textrm{a}_{\bm{k},\omega} }{\partial \omega}\right|_{\omega\to0} - \textrm{h.c.} \right),
	\\ \label{eq:coefficient pi-3'} 
\hat{I}_{0 \xi}^{(2)} 
	& = \frac{1}{2} \sum_{\bm{k}}   \left(  \hat{g}^\textrm{r}_{\bm{k}}  \frac{\partial \hat{g}^\textrm{a}_{\bm{k}} }{\partial k_\xi} - \textrm{h.c.} \right),
	\\ \label{eq:coefficient pi-4'} 
\hat{I}_{0 \xi \xi' }^{(3)} 
	& = \frac{1}{8} \sum_{\bm{k}}   \left(  \hat{g}^\textrm{r}_{\bm{k}}  \frac{\partial^2 \hat{g}^\textrm{a}_{\bm{k}} }{\partial k_\xi \partial k_{\xi'} } 
						+ \frac{\partial^2  \hat{g}^\textrm{r}_{\bm{k}}}{\partial k_\xi \partial k_{\xi'} }  \hat{g}^\textrm{a}_{\bm{k}}
						- 2 \frac{\partial  \hat{g}^\textrm{r}_{\bm{k}}}{\partial k_\xi}   \frac{\partial \hat{g}^\textrm{a}_{\bm{k}} }{\partial k_{\xi'}} 
						 \right).
\end{align}

\subsection{Calculation of $\hat{I}_0^{(0)} $}
We will calculate $\hat{I}_0^{(0)}$ in Eq. (\ref{eq:coefficient pi-1'}).
By using $\hat{g}^\textrm{r}_{\bm{k}}  \hat{g}^\textrm{a}_{\bm{k}} = D^\textrm{r} D^\textrm{a} |\hat{\mathcal{Q}}|^2$ and $\langle \hat{\mathcal{Q}}  \hat{\mathcal{Q}}^\dag \rangle_k 
 =  [  |h|^2  + \hbar^2 \tilde{v}_{\textrm{F}}^2 k^2 ] $, $\hat{I}_0^{(0)}$ becomes
\begin{align} 
\hat{I}^{(0)}_{0}
	& = \sum_{k}   |D^\textrm{r}|^2  [  |h|^2  + \hbar^2 \tilde{v}_{\textrm{F}}^2 k^2 ].
\end{align}
The above equation can be estimated around the Fermi energy, $k \to k_{\textrm{F}} \equiv  \epsilon_{\textrm{F}}/(\hbar \tilde{v}_{\textrm{F}})$ as 
\begin{align}
\hat{I}^{(0)}_{0} & \sim [  |h|^2  + \hbar^2 \tilde{v}_{\textrm{F}}^2 k_{\textrm{F}}^2 ]  \sum_{k}   |D^\textrm{r}|^2  
	  =  \frac{\pi\nu_e }{2\eta }.
\end{align}

\subsection{Calculation of $\hat{I}^{(1)}_{0 }$}
$\hat{I}^{(1)}_{0 }$ in Eq. (\ref{eq:coefficient pi-2'}) is calculated by using 
\begin{align} 
\hat{g}^\textrm{r}_{\bm{k}} \frac{\partial \hat{g}^\textrm{a}_{{\bm{k}},\omega}}{\partial \omega}|_{\omega \to 0}
	& = -\hbar D^\textrm{r} (D^\textrm{a})^2 \hat{\mathcal{Q}}  (\hat{\mathcal{Q}}^\dag)^2, 
	\\ 
|Q|^2  & = \{ |h|^2 + \hbar^2 \tilde{v}_{\textrm{F}}^2 k^2 \} + (h+h^*) \hbar \tilde{v}_{\textrm{F}} \bm{k}\cdot (\bm{z} \times \hat{\bm{\sigma}}), 
	\\
\langle \hat{\mathcal{Q}} (\hat{\mathcal{Q}}^\dag)^2 \rangle_k 
	& =  |h|^2 h^* + (h+2h^*) \hbar^2 \tilde{v}_{\textrm{F}}^2 k^2, 
	\\
\langle \hat{\mathcal{Q}}^2 \hat{\mathcal{Q}}^\dag \rangle_k 
	& =  |h|^2 h + (2 h+h^*) \hbar^2 \tilde{v}_{\textrm{F}}^2 k^2, 
\end{align}
$h+h^*=2\epsilon_{\textrm{F}}$, and $h^* D^\textrm{a} - h D^\textrm{r} = \epsilon_{\textrm{F}} (D^\textrm{a} - D^\textrm{r}) - i\eta(D^\textrm{a} + D^\textrm{r}) $
as 
\begin{align} \notag
\hat{I}^{(1)}_{0}  
	&  =  -\frac{\hbar}{2} \sum_{\bm{k}} 
	 \biggl[    \epsilon_{\textrm{F}} ( |h|^2 + \hbar^2 \tilde{v}_{\textrm{F}}^2 k^2 ) |D^\textrm{r}|^2 (D^\textrm{a} - D^\textrm{r}) 
		\\  \notag 
		& \ \ \ \ \ \ \ \ \ \ \ \ \ \ \ -i\eta ( |h|^2 + \hbar^2 \tilde{v}_{\textrm{F}}^2 k^2 ) |D^\textrm{r}|^2 (D^\textrm{a} + D^\textrm{r}) 
		 + 2 \epsilon_{\textrm{F}}  \hbar^2 \tilde{v}_{\textrm{F}}^2 k^2 |D^\textrm{r}|^2 (D^\textrm{a} - D^\textrm{r}) \biggr]
         \\
         & =  -\frac{i\pi\nu_e }{4\eta^2}\hbar  + o\left(\frac{\hbar}{\epsilon_{\textrm{F}}\tau}\right).
\end{align}

\subsection{Calculation of $\hat{I}^{(2)}_{0\xi }$}
$\hat{I}^{(2)}_{0\xi }$ in Eq. (\ref{eq:coefficient pi-3'}) is obtained by
\begin{align} 
\hat{I}^{(2)}_{0\xi } 
	= &  \frac{\hbar \tilde{v}_{\rm{F}}}{2}\sum_k 
		\epsilon_{\xi z \ell} \hat{\sigma}_\ell \biggl[   (h - h^*)  
				+ 2 \epsilon_{\textrm{F}} \hbar^2 \tilde{v}_{\textrm{F}}^2 k^2  (D^\textrm{a} -D^\textrm{r}) 
		\biggr]|D^\textrm{r}|^2 
	\simeq   \frac{\pi\nu_e }{ 2\eta} \frac{i \tilde{v}_{\textrm{F}} \tau}{2}  \epsilon_{\xi z \ell} \hat{\sigma}_\ell. 
\end{align}
Here, we have used the following equations 
\begin{align} 
\langle  k_\xi |\hat{\mathcal{Q}}|^2  \rangle_k
	& =  \frac{1}{2} \hbar \tilde{v}_{\textrm{F}}  (h+h^*)k^2  (\bm{z} \times \hat{\bm{\sigma}}) _\xi 
	   = \epsilon_{\textrm{F}} \hbar \tilde{v}_{\textrm{F}} k^2 \epsilon_{\xi z \alpha} \hat{\sigma}_\alpha,
	\\ 
 \langle \frac{\partial \hat{g}^\textrm{r}_{\bm{k}}}{\partial k_\xi}  \hat{g}^\textrm{a}_{\bm{k}}  \rangle_k
	& = \hbar \tilde{v}_{\textrm{F}} \epsilon_{\xi z \ell} \hat{\sigma}_\ell \bigl[  h^*  + 2\hbar^2 \tilde{v}_{\textrm{F}}^2 k^2 \epsilon_{\textrm{F}}   D^\textrm{r} \bigr]|D^\textrm{r}|^2.
\end{align}

\subsection{Calculation of $\hat{I}^{(3)}_{0\xi\xi' }$}
$\frac{\partial \hat{g}^\textrm{r}}{\partial k_\alpha} \frac{\partial \hat{g}^\textrm{a}}{\partial k_\beta} $ becomes  
\begin{align} \notag 
\frac{1}{\hbar^2 \tilde{v}_{\textrm{F}}^2 } \frac{\partial \hat{g}^\textrm{r}}{\partial k_\alpha} \frac{\partial \hat{g}^\textrm{a}}{\partial k_\beta}
	= &\epsilon_{\alpha z \ell} \epsilon_{\beta z \ell'} \hat{\sigma}_\ell \hat{\sigma}_{\ell'} |D^\textrm{r}|^2
		+ 4\hbar^2 \tilde{v}_{\textrm{F}}^2 k_\alpha k_\beta |\hat{\mathcal{Q}}|^2 |D^\textrm{r}|^4
		\\
	      & + 2\hbar \tilde{v}_{\textrm{F}}  \epsilon_{\alpha z \ell} \{ \hat{\sigma}_\ell k_\beta  \hat{\mathcal{Q}}^\dag\}  |D^\textrm{r} |^2 D^\textrm{a}   
		 + 2\hbar \tilde{v}_{\textrm{F}}  \epsilon_{\beta z {\ell'}} \{  k_\alpha \hat{\mathcal{Q}} \hat{\sigma}_{\ell'}\}  |D^\textrm{r} |^2 D^\textrm{r} .
\end{align}
The average $\langle  \frac{\partial \hat{g}^\textrm{r}}{\partial k_\alpha} \frac{\partial \hat{g}^\textrm{a}}{\partial k_\beta}  \rangle_k$ is reduced to be
\begin{align} \notag 
\frac{1}{\hbar^2 \tilde{v}_{\textrm{F}}^2 } \langle  \frac{\partial \hat{g}^\textrm{r}}{\partial k_\alpha} \frac{\partial \hat{g}^\textrm{a}}{\partial k_\beta}  \rangle_k
	 = & \epsilon_{\alpha z \ell} \epsilon_{\beta z \ell'} \hat{\sigma}_\ell \hat{\sigma}_{\ell'} |D^\textrm{r}|^2
		+ 2 \delta_{\alpha \beta} |h|^2 (\hbar \tilde{v}_{\textrm{F}} k)^2  |D^\textrm{r}|^4
		+ 2 \delta_{\alpha \beta} (\hbar \tilde{v}_{\textrm{F}} k)^4  |D^\textrm{r}|^4
		\\ 
	&       +  \epsilon_{\alpha z \ell} \epsilon_{\beta u z} \hat{\sigma}_\ell \hat{\sigma}_u  (\hbar \tilde{v}_{\textrm{F}} k)^2 |D^\textrm{r}|^2 D^\textrm{a} 
		 + \epsilon_{\beta z \ell} \epsilon_{\alpha u' z}\hat{ \sigma}_{u'} \hat{\sigma}_\ell (\hbar \tilde{v}_{\textrm{F}} k)^2 |D^\textrm{r}|^2 D^\textrm{r} .
\end{align}
After we integrate the above equation as a function of $\bm{k}$, we obtain 
\begin{align} \notag 
\frac{q_\alpha q_\beta  }{\hbar^2 \tilde{v}_{\textrm{F}}^2 } \sum_{\bm{k}}  \frac{\partial \hat{g}^\textrm{r}}{\partial k_\alpha} \frac{\partial \hat{g}^\textrm{a}}{\partial k_\beta}  
	& = \delta_{\alpha\beta} \sum_{\bm{k}} \biggl\{ |D^\textrm{r}|^2
		+ 2 [ |h|^2 (\hbar \tilde{v}_{\textrm{F}} k)^2 + (\hbar \tilde{v}_{\textrm{F}} k)^4] |D^\textrm{r}|^4 ] 
		 - (\hbar \tilde{v}_{\textrm{F}} k)^2 |D^\textrm{r}|^2 (D^\textrm{a} +D^\textrm{r}) \biggr\}q_\alpha q_\beta
	\\
	& = \frac{\pi\nu_e  }{8\eta^3} q^2   + o\left(\frac{\hbar}{\epsilon_{\textrm{F}}\tau}\right).
\end{align}
Here, we have used 
\begin{align}
q_\alpha q_\beta \epsilon_{\alpha z \ell} \delta_{\beta \ell}  & =0, 
	\\
q_\alpha q_\beta \epsilon_{\alpha z \ell} \epsilon_{\beta z \ell'} \hat{\sigma}_\ell \hat{\sigma}_{\ell'}  &= q_\alpha q_\beta \delta_{\alpha\beta},
	\\ 
q_\alpha q_\beta \epsilon_{\alpha \ell z} \epsilon_{\beta u z} \epsilon_{\ell u \xi} 
	& = q_\alpha q_\beta (\delta_{\alpha \beta } \delta_{\ell u } - \delta_{\alpha u} \delta_{\beta \ell}) \epsilon_{\ell u \xi}  
	 = 0, 
	\\ 
 q_\alpha q_\beta \epsilon_{\alpha z \ell} \epsilon_{\beta u z}
	( \hat{\sigma}_\ell \hat{\sigma}_u  D^\textrm{a}  + \hat{\sigma}_u \hat{\sigma}_\ell  D^\textrm{r} )
& = - q_\alpha q_\beta
	 \delta_{\alpha \beta } (D^\textrm{a} + D^\textrm{r}),
\end{align}
Thus, $q_\xi q_{\xi'} I^{(3)}_{0\xi \xi' }$ becomes 
\begin{align}
q_\xi q_{\xi'} I^{(2)}_{0\xi \xi' }
	& = - \hbar^2 \tilde{v}_{\textrm{F}}^2 \frac{\pi\nu_e  }{16\eta^3} q^2  + o\left(\frac{\hbar}{\epsilon_{\textrm{F}}\tau}\right).
\end{align}

\section{Calculation of $\hat{\Gamma}^{\textrm{rr}}_{\nu} $  \label{sec:A-C} }
We estimate $\hat{\Gamma}^{\textrm{rr}}_{\nu}= n_{\textrm{i}} u_{\textrm{i}}^2\sum_k
\hat{g}^\textrm{r}_{\bm{k}-\bm{\frac{q}{2}},\omega-\frac{\Omega}{2}} \hat{\sigma}_\mu \hat{g}^\textrm{r}_{\bm{k}+\bm{\frac{q}{2}},\omega+\frac{\Omega}{2}}$ by using the same formalism in Appendices \ref{sec:A} and \ref{sec:B}.
Here, $\hat{\Gamma}^{\textrm{rr}}_{\nu} = n_\textrm{i}u_\textrm{i}^2\sum_k \hat{g}^\textrm{r}_{\bm{k}-\bm{\frac{q}{2}},\omega-\frac{\Omega}{2}}
\hat{\sigma}_\mu \hat{g}^\textrm{r}_{\bm{k}+\bm{\frac{q}{2}},\omega+\frac{\Omega}{2}}$ can be expanded as
\begin{align}
n_\textrm{i}u_\textrm{i}^2
\sum_k
\hat{g}^\textrm{r}_{\bm{k}-\bm{\frac{q}{2}},\omega-\frac{\Omega}{2}}
\hat{\sigma}_\mu
\hat{g}^\textrm{r}_{\bm{k}+\bm{\frac{q}{2}},\omega+\frac{\Omega}{2}}
=&\hat{C}^{\textrm{rr}(0)}_{\mu}+\Omega \hat{C}^{\textrm{rr}(1)}_{\mu}+\sum_{\xi=x,y}q_\xi \hat{C}^{\textrm{rr}(2)}_{\mu\xi}+\sum_{\xi,\xi'=x,y}q_\xi q_{\xi'} \hat{C}^{\textrm{rr}(3)}_{\mu\xi\xi'} +O(q^3, q\Omega, \Omega^2),
\end{align}
where coefficients in the above equation are given by 
\begin{align}
\hat{C}^{\textrm{rr}(0)}_{\mu}\label{eq:Crr0}
	=&n_\textrm{i}u_\textrm{i}^2\sum_{k}\left[\hat{g}^\textrm{r}\hat{\sigma}_\mu \hat{g}^\textrm{r}\right] =\left\{
\begin{array}{ll}
\displaystyle-\frac{n_\textrm{i}u_\textrm{i}^2\nu_e}{2\epsilon_F}\hat{\sigma}_\mu & (\mu=x,y)\\
\displaystyle- \frac{in_\textrm{i}u_\textrm{i}^2\pi\nu_e}{2\epsilon_F}\arg(-\hbar\omega-\epsilon_\textrm{F}+i\eta)\hat{\sigma}_z & (\mu=z)\\
\displaystyle-\frac{n_\textrm{i}u_\textrm{i}^2\nu_e}{\epsilon_F} + \frac{in_\textrm{i}u_\textrm{i}^2\pi\nu_e}{2\epsilon_F}\arg(-\hbar\omega-\epsilon_\textrm{F}+i\eta) & (\mu=0)
\end{array}\right., 
	\\
\hat{C}^{\textrm{rr}(1)}_{\mu}\label{eq:Crr1}
	=&\frac{n_\textrm{i}u_\textrm{i}^2}{2}\sum_{ k}\left[ \hat{g}^\textrm{r}\hat{\sigma}_\mu\frac{\partial \hat{g}^\textrm{r}}{\partial \omega}
-\frac{\partial \hat{g}^\textrm{r}}{\partial \omega}\hat{\sigma}_\mu \hat{g}^\textrm{r} \right]
        = 0,
	 \\
\hat{C}^{\textrm{rr}(2)}_{\mu\xi}\label{eq:Crr2}
	=&\frac{n_\textrm{i}u_\textrm{i}^2}{2}\sum_{k}\left[\hat{g}^\textrm{r}\hat{\sigma}_\mu\frac{\partial \hat{g}^\textrm{r}}{\partial k_\xi}
-\frac{\partial \hat{g}^\textrm{r}}{\partial k_\xi}\hat{\sigma}_\mu \hat{g}^\textrm{r} \right]
	=\left\{
		\begin{array}{ll}
		\displaystyle\frac{n_\textrm{i}u_\textrm{i}^2\hbar v_F\nu_e}{2\epsilon_F[(\hbar\omega+\epsilon_F)^2+\eta^2]}[i(\hbar\omega+\epsilon_F)+\eta]\delta_{\xi \mu}	\hat{\sigma}_z& (\mu=x,y)
			\\
		\displaystyle-\frac{n_\textrm{i}u_\textrm{i}^2\hbar v_F\nu_e}{2\epsilon_F[(\hbar\omega+\epsilon_F)^2+\eta^2]}[i(\hbar\omega+\epsilon_F)+\eta]\hat{\sigma}_\xi& (\mu=z)
			\\
		0 & (\mu=0)
		\end{array}\right., 
	\\ \notag 
\hat{C}^{\textrm{rr}(3)}_{\mu\xi\xi'}
	=&\frac{n_\textrm{i}u_\textrm{i}^2}{4}\sum_{k}\left[\hat{g}^\textrm{r}\hat{\sigma}_\mu\frac{\partial^2 \hat{g}^\textrm{r}}{\partial k_\xi\partial k_{\xi'}}
+\frac{\partial^2 \hat{g}^\textrm{r} }{\partial k_\xi\partial k_{\xi'}}\hat{\sigma}_\mu \hat{g}^\textrm{r}\right]
	\\\label{eq:Crr3}
	=&\left\{
\begin{array}{ll}
\displaystyle\frac{n_\textrm{i}u_\textrm{i}^2\hbar^2v_F^2\nu_e}{12\epsilon_F[(\hbar\omega+\epsilon_F)^2+\eta^2]^2}\left[(\hbar\omega+\epsilon_{\textrm{F}})^2-\eta^2-2i\eta(\hbar\omega+\epsilon_{\textrm{F}})\right]&\\
	\displaystyle\hspace{20mm}\times[\epsilon_{\xi z \mu}(\bm{z}\times\hat{\bm{\sigma}})_{\xi'}+\epsilon_{\xi' z \mu}(\bm{z}\times\hat{\bm{\sigma}})_\xi -2\delta_{\xi\xi'}\hat{\sigma}_\mu]&(\mu=x,y)\\
\displaystyle-\frac{n_\textrm{i}u_\textrm{i}^2\hbar^2v_F^2\nu_e}{4\epsilon_F[(\hbar\omega+\epsilon_F)^2+\eta^2]^2}\left[(\hbar\omega+\epsilon_{\textrm{F}})^2-\eta^2-2i\eta(\hbar\omega+\epsilon_{\textrm{F}})\right]\delta_{\xi\xi'}\hat{\sigma}_z&(\mu=z)\\
\displaystyle\frac{n_\textrm{i}u_\textrm{i}^2\hbar^2v_F^2\nu_e}{12\epsilon_F[(\hbar\omega+\epsilon_F)^2+\eta^2]^2}\left[(\hbar\omega+\epsilon_{\textrm{F}})^2-\eta^2-2i\eta(\hbar\omega+\epsilon_{\textrm{F}})\right]\delta_{\xi\xi'}\hat{\sigma}_0&(\mu=0)
\end{array}\right. .
\end{align}
From the Eqs. (\ref{eq:Crr0})-(\ref{eq:Crr3}) and $n_{\textrm{i}} u_{\textrm{i}}^2 \pi\nu_e/ \eta=2 $, the elements of $\hat{\Gamma}^{\textrm{rr}}_{\nu} $ 
is negligibly small as compared with the ones of $\hat{\Gamma}^{\textrm{ra}}_{\nu} $, since $ \frac{\hbar}{\epsilon_{\textrm{F}}\tau} \ll 1$ is satisfied.

\section{Calculation of $\hat{\Pi}^{\textrm{rr}}_{\nu} + \hat{\Pi}^{\textrm{aa}}_{\nu} \ \ (\nu=x,y)$  \label{sec:D} }
We estimate the response function composed of only the retarded (advanced) Green's functions $\hat{\Pi}^{\textrm{rr}}_{\nu}(\bm{q}, \Omega)$ ($\hat{\Pi}^{\textrm{aa}}_{\nu}(\bm{q}, \Omega)$) using the same formalism in Appendices \ref{sec:A}, \ref{sec:B} and \ref{sec:A-C}.
From Eqs. (\ref{eq:3-11}) and (\ref{eq:3-11-2}), $\hat{\Pi}^{\textrm{rr}}_{\nu}(\bm{q}, \Omega) + \hat{\Pi}^{\textrm{aa}}_{\nu}(\bm{q}, \Omega)$ are written as
\begin{align} \notag 
\hat{\Pi}^{\textrm{rr}}_{ \nu} (\bm{q}, \Omega) + \hat{\Pi}^{\textrm{aa}}_{\nu}(\bm{q}, \Omega)
	& = - \sum_{\bm{k}, \omega} \biggl\{ f_{\omega}   \left[\hat{g}^\textrm{r}_{\bm{k}-\frac{\bm{q}}{2},\omega-\frac{\Omega}{2}} \hat{\sigma}_\nu \hat{g}^\textrm{r}_{\bm{k}+\frac{\bm{q}}{2},\omega+\frac{\Omega}{2}}
		- \left(\hat{g}^\textrm{r}_{\bm{k}+\frac{\bm{q}}{2},\omega+\frac{\Omega}{2}} \hat{\sigma}_\nu \hat{g}^\textrm{r}_{\bm{k}-\frac{\bm{q}}{2},\omega-\frac{\Omega}{2}}\right)^\dag\right] 
	\\&\hspace{20mm}		
		+ \frac{1}{2} \Omega f'_{\omega} \biggl[   \hat{g}^\textrm{r}_{\bm{k}-\frac{\bm{q}}{2},\omega-\frac{\Omega}{2}} \hat{\sigma}_\nu  \hat{g}^\textrm{r}_{\bm{k}+\frac{\bm{q}}{2},\omega+\frac{\Omega}{2}} 
		+  \left(\hat{g}^\textrm{r}_{\bm{k}+\frac{\bm{q}}{2},\omega+\frac{\Omega}{2}} \hat{\sigma}_\nu \hat{g}^\textrm{r}_{\bm{k}-\frac{\bm{q}}{2},\omega-\frac{\Omega}{2}}\right)^\dag \biggr]\biggr\}.
\label{eq:Pirr+aa}
\end{align}
The magnitude of second term of the above equation is smaller than that of $\hat{\Pi}^{\textrm{ra}}_{\nu}$ [see Eqs. (\ref{eq:3-10}) and Appendices \ref{sec:A}, \ref{sec:B} and \ref{sec:A-C}  ].
The first term is expanded as
\begin{align}\notag
&- \sum_{\bm{k}, \omega} f_{\omega}   \left[\hat{g}^\textrm{r}_{\bm{k}-\frac{\bm{q}}{2},\omega-\frac{\Omega}{2}} \hat{\sigma}_\nu \hat{g}^\textrm{r}_{\bm{k}+\frac{\bm{q}}{2},\omega+\frac{\Omega}{2}}
		- \left(\hat{g}^\textrm{r}_{\bm{k}+\frac{\bm{q}}{2},\omega+\frac{\Omega}{2}} \hat{\sigma}_\nu \hat{g}^\textrm{r}_{\bm{k}-\frac{\bm{q}}{2},\omega-\frac{\Omega}{2}}\right)^\dag\right] 
	\\
	 = & \hat{D}^{(0)}_{\nu}+\Omega \hat{D}^{(1)}_{\nu}+\sum_{\xi=x,y}q_\xi \hat{D}^{(2)}_{\nu\xi}+O(q^2, q\Omega, \Omega^2),
\end{align}
where $\hat{D}^{(0)}_{\nu}$, $\hat{D}^{(1)}_{\nu}$ and $\hat{D}^{(0)}_{\nu\xi}$ in the above equation are given by 
\begin{align}\label{eq:D0nu}
\hat{D}^{(0)}_{\nu}
	=&-\sum_{\bm{k}, \omega}f_{\omega}\left[\hat{g}_{\bm{k},\omega}^\textrm{r}\hat{\sigma}_\nu \hat{g}_{\bm{k},\omega}^\textrm{r} - \textrm{h.c.} \right] = 0,
	\\\label{eq:D1nu}
\hat{D}^{(1)}_{\nu}
	=&-\frac{1}{2}\sum_{\bm{k}, \omega}\left[ \left(\hat{g}_{\bm{k},\omega}^\textrm{r}\hat{\sigma}_\mu\frac{\partial \hat{g}_{\bm{k},\omega}^\textrm{r}}{\partial \omega}
-\frac{\partial \hat{g}_{\bm{k},\omega}^\textrm{r}}{\partial \omega}\hat{\sigma}_\mu \hat{g}_{\bm{k},\omega}^\textrm{r}\right) + \textrm{h.c.} \right]
        = 0,
	\\\label{eq:D2nu}
\hat{D}^{(2)}_{\nu\xi}
	=&-\frac{1}{2}\sum_{\bm{k}, \omega}f_{\omega}\left[\left(\hat{g}_{\bm{k},\omega}^\textrm{r}\hat{\sigma}_\mu\frac{\partial \hat{g}_{\bm{k},\omega}^\textrm{r}}{\partial k_\xi}
-\frac{\partial \hat{g}_{\bm{k},\omega}^\textrm{r}}{\partial k_\xi}\hat{\sigma}_\mu \hat{g}_{\bm{k},\omega}^\textrm{r}\right) + \textrm{h.c.} \right]
	= -\frac{\pi \tilde{v}_{\textrm{F}}\nu_{\textrm{e}}}{2\epsilon_{\textrm{F}}}\hat{\sigma}_z + o\left(\frac{\hbar}{\epsilon_{\textrm{F}}\tau}\right).
\end{align}
Here, we have used $f_\omega=\theta(-\omega)$ in the above equation, where $\theta(x)$ is a step function. 
From Eqs. (\ref{eq:3-2}), (\ref{eq:3-6}), (\ref{eq:3-33}) and (\ref{eq:D2nu}), there are spin density and charge current induced by the magnetic field $(i\bm{q}\times\bm{A}_{\textrm{em}})$. 
Since we consider only the electric field, we ignore the densities induced by the magnetic field.

We find that the order of $(\hat{\Pi}^{\textrm{rr}}_{\nu} + \hat{\Pi}^{\textrm{aa}}_{\nu})/\hat{\Pi}^{\textrm{ra}}_{\nu}$  are $\hbar/\epsilon_{\textrm{F}}\tau$ and $\hat{\Pi}^{\textrm{rr}}_{\nu} + \hat{\Pi}^{\textrm{aa}}_{\nu}$ are negligibly small as compared with $\hat{\Pi}^{\textrm{ra}}_{\nu}$.

\section{Calculation of integral}\label{sec:E}
We will show the following integrals as a functions of $\bm{k}$. 
The integrals are obtained by  
\begin{align} 
\sum_{k}   D^\textrm{r} (D^\textrm{a})^2 
		& \simeq \frac{\nu_e}{16\eta^2 \epsilon_{\textrm{F}}^3} \left( i\pi - \frac{4\eta^2}{\epsilon_{\textrm{F}}^2 }\right),
	\\ 
\sum_{k}   \hbar^2 \tilde{v}_{\textrm{F}}^2 k^2 D^\textrm{r} (D^\textrm{a})^2  
		& \simeq \frac{ i\pi \nu_e }{16\eta^2 \epsilon_{\textrm{F}}} \left( 1+ i \frac{2\eta}{\epsilon_{\textrm{F}}}\right), 
	\\\label{eq:D4}
\sum_{k} [ (h^*)^2 + \hbar^2 \tilde{v}_{\textrm{F}}^2 k^2]   D^\textrm{r} (D^\textrm{a})^2 
	& \simeq  \frac{\nu_e}{8\eta^2 \epsilon_{\textrm{F}}} 
		\biggl[ i\pi - \frac{2\eta^2}{\epsilon_{\textrm{F}}^2 }  \biggr], 
	\\\label{eq:D5}
\sum_{k}  [  h(h^*)^2 + (h - 2h^*)\hbar^2 \tilde{v}_{\textrm{F}}^2 k^2 ] D^\textrm{r} (D^\textrm{a})^2 
	& \simeq \frac{-i\pi\nu_e}{4\epsilon_\textrm{F}^2}\left(1+i\frac{\eta}{\epsilon_{\textrm{F}}} \right),
	\\
\sum_k \hbar^2\tilde{v}_{\textrm{F}}^2 k^2  |D^\textrm{r}|^2 ( D^\textrm{a} - D^\textrm{r})
	& \simeq \frac{ i\pi \nu_e }{8\eta^2 \epsilon_{\textrm{F}}}, 
	\\
\sum_k  |D^\textrm{r}|^2 
	& \simeq \frac{\pi\nu_e}{4\eta \epsilon_{\textrm{F}}^2},
	\\
\sum_k \bigl[ (D^\textrm{r})^2 D^\textrm{a} + (D^\textrm{a})^2 D^\textrm{r} \bigr] 
	& \simeq  -\frac{\nu_e}{2 \epsilon_{\textrm{F}}^5}, 
	\\
\sum_k (\hbar \tilde{v}_{\textrm{F}} k )^2 \bigl[ (D^\textrm{r})^2 D^\textrm{a} + (D^\textrm{a})^2 D^\textrm{r} \bigr] 
	& \simeq - \frac{ \pi \nu_e }{4 \eta \epsilon_{\textrm{F}}^2}, 
	\\ 
\sum_k (\hbar \tilde{v}_{\textrm{F}} k )^2 (D^\textrm{r})^3 D^\textrm{a} 
	& \simeq  -\frac{\pi \nu_e}{64 \epsilon_{\textrm{F}}^2 \eta^3} \left( 1- i\frac{2\eta}{\epsilon_{\textrm{F}}} \right), 
	 \\
\sum_k (\hbar \tilde{v}_{\textrm{F}} k )^2 \{ (D^\textrm{r})^3 D^\textrm{a} + (D^\textrm{a})^3 D^\textrm{r}\}
	& \simeq -\frac{\pi \nu_e}{32 \epsilon_{\textrm{F}}^2 \eta^3},
	\\
\sum_k (\hbar \tilde{v}_{\textrm{F}} k )^4 (D^\textrm{r})^3 D^\textrm{a}
	& \simeq -\frac{\pi \nu_e }{32 \eta^3}, 
	\\
\sum_k(\hbar v_Fk)^2(D^\textrm{r})^{n}
	&\simeq-\frac{1}{n-1}\sum_k(D^\textrm{r})^{n-1} \ \ \ \ \ (n\geq3),
	\\
\sum_k(\hbar v_Fk)^4(D^\textrm{r})^{n} & =\frac{2}{(n-1)(n-2)}\sum_k(D^\textrm{r})^{n-2} \ \ \ \ \ (n\geq4),
	\\
\sum_k(\hbar v_Fk)^2(D^\textrm{r})^{3}
	&\simeq-\frac{1}{2}\sum_k(D^\textrm{r})^{2},
	\\
\sum_k(\hbar v_Fk)^2(D^\textrm{r})^{4}
	&\simeq -\frac{1}{3}\sum_k(D^\textrm{r})^{3}= \frac{\pi\nu_e }{32\epsilon_{\textrm{F}}^2 \eta^3},
	\\
\sum_k(\hbar v_Fk)^4(D^\textrm{r})^{4}
	&\simeq \frac{1}{3}\sum_k(D^\textrm{r})^{2} = \frac{\pi\nu_e }{32 \eta^3}.
\end{align}
where $\sum_{\bm{k}}$ is defined by  
\begin{align} 
\sum_{\bm{k}} &\equiv  \frac{1}{(2\pi)^2} \int_0^{2\pi} d\theta \int_0^\infty  k dk
	 =  \frac{\nu_e}{2\pi \epsilon_{\textrm{F}}} \int_0^{2\pi} d\theta \int_0^\infty  \epsilon d\epsilon
	 =  \frac{\nu_e}{4\pi \epsilon_{\textrm{F}}} \int_0^{2\pi} d\theta \int_0^\infty  dx.
\end{align}
Here, we have used $(D^\textrm{r})^{n} = \frac{1}{2(n-1)\hbar^2v_F^2k_\xi}\frac{\partial (D^\textrm{r})^{n-1}}{\partial k_\xi}$ in the above equation.

\section{Charge conservation}\label{sec:F} 
To check validity of our results, we substitute the charge current and charge density in Eqs. (\ref{eq:3-32}) and (\ref{eq:3-24}) into the charge conservation law $\dot{\rho}_e + \bm{\nabla}\cdot \bm{j} =0$. 
From Eq. (\ref{eq:3-23}), $\dot{\rho}_e $ becomes 
\begin{align} 
\dot{\rho}_e 
	& =  \frac{  e^2  \tilde{v}_{\textrm{F}}^2 \ell  \nu_e }{L^2}  \sum_{\bm{q},\Omega}e^{i[\Omega t-\bm{q}\cdot\bm{x}]} 
		\frac{ i \Omega^2 q_\nu  }{q^2 \ell^2 + i\Omega \tau} A_{\textrm{em},\nu} .
\end{align}
From Eq. (\ref{eq:3-21}), (\ref{eq:3-22}) and $\bm{j} =  2 e \tilde{v}_\textrm{F}  (\bm{z} \times \bm{s})$, $\bm{\nabla}\cdot \bm{j}$ becomes
\begin{align}
\nabla_x j_{x}	
	& = \frac{e^2 \tilde{v}_{\textrm{F}}^2 \nu_e \tau}{ L^2} 
		  \sum_{\bm{q},\Omega}e^{i[\Omega t-\bm{q}\cdot\bm{x}]}
		   \biggl[  - \Omega q_x A_{{\textrm{em}}, x} + \biggl\{ \frac{ \Omega q_x^2 q_y \ell^2 }{q^2 \ell^2 + i\Omega \tau} A_{{\textrm{em}}, y} + \frac{ \Omega q_x^3 \ell^2 }{q^2 \ell^2 + i\Omega \tau} A_{{\textrm{em}}, x}  \biggr\} \biggr],
		   \\
\nabla_y j_{y}	& = \frac{e^2 \tilde{v}_{\textrm{F}}^2 \nu_e \tau}{ L^2} 
		  \sum_{\bm{q},\Omega}e^{i[\Omega t-\bm{q}\cdot\bm{x}]}
		   \biggl[  -\Omega q_y A_{{\textrm{em}}, y} + \biggl\{ \frac{ \Omega q_x q_y^2 \ell^2 }{q^2 \ell^2 + i\Omega \tau} A_{{\textrm{em}}, x} +  \frac{\Omega q_y^3 \ell^2 }{q^2 \ell^2 + i\Omega \tau} A_{{\textrm{em}}, y}  \biggr\} \biggr],
	 \\ 
\nabla_x j_x + \nabla_y j_y   
	&=   -\frac{ e^2 \tilde{v}_{\textrm{F}}^2 \ell \nu_e }{ L^2} 
		  \sum_{\bm{q},\Omega}e^{i[\Omega t-\bm{q}\cdot\bm{x}]}
		    \frac{ i \Omega^2  q_\nu 	   }{q^2 \ell^2 + i\Omega \tau}    A_{{\textrm{em}}, \nu}.
\end{align}
Therefore, $\rho_e$ and $\bm{j}$ follow the charge conservation law, $\dot{\rho}_e + \bm{\nabla}\cdot \bm{j} = 0$.

\section{Diffusive Green's function $\mathcal{D}$}\label{sec:G} 
Diffusive Green's function on the disordered surface of the TI can be integrated as follows: 
\begin{align} \notag 
\sum_\Omega \frac{ e^{i(\Omega t -\bm{q}\cdot\bm{x}) }}{i\Omega + 2Dq^2}
	& \sim \frac{1}{2\pi i} \int_{-\infty}^{\infty}	\frac{d\Omega}{\Omega -i 2Dq^2}
	\\
	& = \theta(t) \exp{[-2Dt q^2 -i\bm{q}\cdot\bm{x}] }, 
	\\ \notag 
\sum_{q_x} e^{-(2D t q_x^2 + i q_x x)} 
	& \sim \frac{1}{2\pi}\int_{-\infty}^\infty dq_x e^{-(2D t q_x^2 + i q_x x)}
	\\
	& = \frac{\sqrt{\pi}}{2\pi \sqrt{2Dt}} \exp{\left(- \frac{2x^2}{Dt}\right)}.
\end{align}
Thus, from the above equations, $\mathcal{D}$ in the coordinates space is obtained by 
\begin{align}
\mathcal{D}(\bm{x},t) 
	& \sim  \frac{\theta (t)}{8\pi Dt} \exp{\left[- \frac{1}{8Dt} (x^2+y^2)\right]}.
\end{align}




\begin{references}
\bibitem{rf:estimation}Using $\frac{\hbar}{2\tau}=\frac{1}{2}\pi\nu_e n_i u_0^2$, $\nu_e = \frac{\epsilon_\textrm{F}}{2\pi \hbar^2 \tilde{v}^2_\textrm{F}}$, and $\tilde{v}_\textrm{F}/v_\textrm{F}=(1+\xi)^{-1}$, 
we can obtain the relation: $\xi (1+\xi)^2 = \frac{\hbar}{4\pi \epsilon_\textrm{F}\tau}$.
Because of $\frac{\hbar}{\epsilon_\textrm{F}\tau}\ll1$, $\xi < \frac{\hbar}{2\pi \epsilon_\textrm{F}\tau}$ becomes negligible small as $\tilde{v}_\textrm{F}/v_\textrm{F} = 1 + o(\frac{\hbar}{\epsilon_\textrm{F}\tau})$.
Streakily speaking, there is no problem to take $\tilde{v}_\textrm{F} \simeq v_\textrm{F}$.
\end{references}
\end{document}